\documentclass[twoside]{article}
\usepackage{a4wide,epsf,cite}
\usepackage[hang,sc,small]{caption}

\begin{document}
\sloppy
%
\newcommand{\figI}      {mt_tb0}                
\newcommand{\figII}     {m1_m2}                 
\newcommand{\figIII}    {chi2_gyebdl}           
\newcommand{\figIV}     {spectrum}              
\newcommand{\figV}      {bsg}                   
\newcommand{\figVI}     {relic}                 
\newcommand{\figVII}    {mchar1}                
\newcommand{\figVIII}   {chi2_gyel}             
\newcommand{\figIX}     {dmb}                   
\newcommand{\figX}      {mt_tb1}                
\newcommand{\figXa}     {running_At}
\newcommand{\figXI}     {mglui}                 
\newcommand{\figXII}    {mu_mglu}               
\newcommand{\figXIII}   {atop_mglu}             
\newcommand{\figXIV}    {abot_mglu}             
\newcommand{\figXV}     {atau_mglu}             
\newcommand{\figXVI}    {mu}                    
\newcommand{\figXVII}   {mh}                    
\newcommand{\figXVIII}  {ma}                    
\newcommand{\figXIXa}   {m1}                    
\newcommand{\figXIXb}   {m2}                    
\newcommand{\figXIXc}   {S1_m1}                 
\newcommand{\figXIXd}   {S2_m2}                 
\newcommand{\figXIXe}   {dmz}                   
\newcommand{\figXX}     {43a_hz0}               
\newcommand{\figXXI}    {49a_hz0-300}           
\newcommand{\figXXII}   {x_hz0_low_high}        
\newcommand{\figXXIII}  {49a_chichi-300}        

\newcommand{\unify}  {plot1}           
\newcommand{\runma}  {plot5}           
\newcommand{\proton} {proton}          
\newcommand{\mbvsmt} {plot3}           
\newcommand{\mhvsmt} {plot7}           
\newcommand{\chisq}  {chi2}            
\newcommand{\chisqp} {chi2prot}        
\newcommand{\mufit}  {mu}              
\newcommand{\mucormh}{plot4}           
%
\newcommand{\bq}{\begin{equation}}
\newcommand{\eq}{\end{equation}}
\newcommand{\beq}  {\begin{eqnarray}}
\newcommand{\eeq}  {\end{eqnarray}}
\newcommand{\rG}   {{\rm GUT} }
\newcommand{\MG}   {{\ifmmode M_\rG         \else $M_\rG$          \fi}}
\newcommand{\mb}   {{\ifmmode m_{b}         \else $m_{b}$          \fi}}
\newcommand{\mt}   {{\ifmmode m_{t}         \else $m_{t}$          \fi}}
\newcommand{\agut} {{\ifmmode \alpha_\rG    \else $\alpha_\rG$     \fi}}
\newcommand{\mgut} {{\ifmmode M_\rG         \else $M_\rG$
    \fi}}
\newcommand{\rPL}  {{\rm Planck}}
\newcommand{\mplanck} {{\ifmmode M_\rPL         \else $M_\rPL$          \fi}}
\newcommand{\rST}  {{\rm SO(10)}}
\newcommand{\msoten} {{\ifmmode M_\rST         \else $M_\rST$          \fi}}
\newcommand{\mze}  {{\ifmmode m_0           \else $m_0$            \fi}}
\newcommand{\mha}  {{\ifmmode m_{1/2}       \else $m_{1/2}$        \fi}}
\newcommand{\tb}   {{\ifmmode \tan\beta     \else $\tan\beta$      \fi}}
\newcommand{\mz}   {{\ifmmode M_{Z}         \else $M_{Z}$          \fi}}
\newcommand{\ai}   {{\ifmmode \alpha_i      \else $\alpha_i$       \fi}}
\newcommand{\aii}  {{\ifmmode \alpha_i^{-1} \else $\alpha_i^{-1}$  \fi}}
\newcommand{\DRbar}{{\ifmmode \overline{DR} \else $ \overline{DR}$ \fi}}
\newcommand{\msusy}{{\ifmmode M_{SUSY}      \else $M_{SUSY}$       \fi}}
\newcommand{\as}   {{\ifmmode \alpha_s      \else $\alpha_s$       \fi}}
\newcommand{\asmz} {{\ifmmode \alpha_s(M_Z) \else $\alpha_s(M_Z)$  \fi}}
\newcommand{\tal}  {{\ifmmode \tilde{\alpha} \else $\tilde{\alpha}$ \fi}}
\newcommand{\rb}[1]{\raisebox{1.5ex}[-1.5ex]{#1}}
\newcommand {\tabs}[1]{\multicolumn{1}{c}{\mbox{\hspace{#1}}}}
\newcommand{\sws}  {{\ifmmode \;\sin^2\theta_W
                     \else    $\;\sin^{2}\theta_{W}$               \fi}}
\newcommand{\cws}  {{\ifmmode \;\cos^2\theta_W  
                     \else    $\;\cos^{2}\theta_{W}$               \fi}}
\newcommand{\sw}   {{\ifmmode\;\sin\theta_W\else $\sin\theta_{W}$  \fi}}
\newcommand{\cw}   {{\ifmmode\;\cos\theta_W\else $\;\cos\theta_{W}$\fi}}
\newcommand{\tw}   {{\ifmmode\;\tan\theta_W\else $\;\tan\theta_{W}$\fi}}
\newcommand{\bsg}  {{\ifmmode b\rightarrow s\gamma
                     \else $b\rightarrow s\gamma$ \fi}}
\newcommand{\Bbsg}  {{\ifmmode BR(\b\rightarrow s\gamma)
\else $BR(b\rightarrow s\gamma)$ \fi}}
\newcommand{\smas}[2]{\tilde{m}^#2_{#1}}
\newcommand{\nn}   {\nonumber \\}
\renewcommand{\floatpagefraction}{0.005}

\newcommand{\Zto}   {\mbox{$\mathrm Z \to$}}

\def\NPB#1#2#3{{\it Nucl.~Phys.} {\bf{B#1}} (19#2) #3}
\def\PLB#1#2#3{{\it Phys.~Lett.} {\bf{B#1}} (19#2) #3}
\def\PRD#1#2#3{{\it Phys.~Rev.} {\bf{D#1}} (19#2) #3}
\def\PRL#1#2#3{{\it Phys.~Rev.~Lett.} {\bf{#1}} (19#2) #3}
\def\ZPC#1#2#3{{\it Z.~Phys.} {\bf C#1} (19#2) #3}
\def\PTP#1#2#3{{\it Prog.~Theor.~Phys.} {\bf#1}  (19#2) #3}
\def\MPL#1#2#3{{\it Mod.~Phys.~Lett.} {\bf#1} (19#2) #3}
\def\PR#1#2#3{{\it Phys.~Rep.} {\bf#1} (19#2) #3}
\def\RMP#1#2#3{{\it Rev.~Mod.~Phys.} {\bf#1} (19#2) #3}
\def\HPA#1#2#3{{\it Helv.~Phys.~Acta} {\bf#1} (19#2) #3}
\def\NIMA#1#2#3{{\it Nucl.~Instr.~and~Meth.} {\bf#1} (19#2) #3}

\begin{titlepage}
\begin{flushright}
\vspace*{-2.2cm}
\noindent
 \hfill IEKP-KA/96-04    \\
 \hfill hep-ph/9603350   \\
 \hfill March 18th, 1996 \\
\end{flushright}
\vspace{1.7cm}
\begin{center} {\Large\bf Combined Fit of Low Energy Constraints \\
         to  Minimal Supersymmetry \\
         and Discovery Potential at LEP II\\}
\vspace{0.5cm}
{\small {\bf  W.~de Boer\footnote{Email: wim.de.boer@cern.ch}, 
 G.~Burkart\footnote{E-mail: gerd@ekp.physik.uni-karlsruhe.de}, 
 R.~Ehret\footnote{E-mail: ralf.ehret@cern.ch}, \\
 J.~Lautenbacher\footnote{E-mail: jens@ekp.physik.uni-karlsruhe.de},
 W.~Oberschulte-Beckmann\footnote{E-mail: wulf@ekp.physik.uni-karlsruhe.de},
 U.~Schwickerath\footnote{E-mail: ulrich@ekp.physik.uni-karlsruhe.de}
\\
}
{\it Inst.\ f\"ur Experimentelle Kernphysik, Univ.\ of Karlsruhe, \\}
{\it Postfach 6980, D-76128 Karlsruhe, Germany  \\} and \\
{\bf V. Bednyakov, D.I. Kazakov\footnote{E-mail: 
kazakovd@thsun1.jinr.dubna.su}, S.G. Kovalenko\footnote{E-mail:
kovalen@lnpnw1.jinr.dubna.su}\\}
{\it Bogoliubov Lab. of Theor. Physics,
     Joint Inst. for Nucl. Research, \\}
{\it 141 980 Dubna, Moscow Region, Russia \\}}
\end{center}

\vspace{0.5cm}

\begin{center}
{\bf Abstract}
\end{center}
\parindent0.0pt \small Within the Constrained Minimal
Supersymmetric Standard Model (CMSSM) it is possible to predict the
low energy gauge couplings and masses of the 3.~generation particles
from a few parameters at the GUT scale. In addition the MSSM predicts
electroweak symmetry breaking due to large radiative corrections from
Yukawa couplings, thus relating the $Z^0$ boson mass to the top quark
mass.

From a $\chi^2$~analysis, in which these constraints can be considered
simultaneously, one can calculate the probability for each point in
the MSGUT parameter space.  The recently measured top quark mass
prefers two solutions for the mixing angle in the Higgs sector:
$\tan\beta$ in the range between 1 and 3 or alternatively
$\tan\beta\approx 25-50$.  For both cases we find a unique $\chi^2$
minimum in the parameter space. From the corresponding most probable
parameters at the GUT scale, the masses of all predicted particles can
be calculated at low energies using the RGE, albeit with rather large
errors due to the logarithmic nature of the running of the masses and
coupling constants.  Our fits include full second order corrections
for the gauge and Yukawa couplings, low energy threshold effects,
contributions of all (s)particles to the Higgs potential and
corrections to $m_b$ from gluinos and higgsinos, which exclude (in our
notation) positive values of the mixing parameter $\mu$ in the Higgs
potential for the large $\tan\beta$ region.

Further constraints can be derived from the branching ratio for the
radiative (penguin) decay of the $b$-quark into $s\gamma$ and the lower
limit on the lifetime of the universe, which requires the dark matter
density due to the Lightest Supersymmetric Particle (LSP) not to
overclose the universe.
 
For the low $\tan\beta$ solution these additional constraints can be
fulfilled simultaneously for quite a large region of the parameter
space. In contrast, for the high $\tan\beta$ solution the correct
value for the $ b \rightarrow s\gamma$ rate is obtained only for small
values of the gaugino scale and electroweak symmetry breaking is
difficult, unless one assumes the minimal SU(5) to be a subgroup of a
larger symmetry group, which is broken between the Planck scale and
the unification scale.  In this case small splittings in the Yukawa
couplings are expected at the unification scale and electroweak
symmetry breaking is easily obtained, provided the Yukawa coupling for
the top quark is slightly above the one for the bottom quark, as
expected e.g. if the larger symmetry group would be SO(10).

For particles, which are most likely to have masses in the LEP II
energy range, the cross sections are given for the various energy
scenarios at LEP II. For low $\tb$ the production of the lightest
Higgs boson, which is expected to have a mass below 103 GeV, is the
most promising channel, while for large $\tb$ the production of
charginos and/or neutralinos covers the preferred parameter space.
\end{titlepage}

\cleardoublepage

\section{Introduction}

Grand Unified Theories (GUT's) in which the electroweak and strong
forces are unified at a scale \MG of the order $10^{16}$ GeV are
strongly constrained by low energy data, if one imposes unification of
gauge- and Yukawa couplings as well as electroweak symmetry breaking.
The Minimal Supersymmetric Standard Model (MSSM) \cite{su5susy} has
become the leading candidate for a GUT after the precisely measured
coupling constants at LEP excluded unification in the Standard Model
\cite{ekn2,abf,lalu}.  In the MSSM the quadratic divergences in the
higher order radiative corrections largely cancel, so one can
calculate the corrections reliably even over many orders of magnitude.
The large hierarchy between the electroweak scale and the unification
scale as well as the different strengths of the forces at low energy
are naturally explained by the radiative corrections\cite{rev}.  Low
energy data on masses and couplings provide strong constraints on the
MSSM parameter space, as discussed recently by many groups
~\cite{rrb92,cpw,bbo,op,chan,langpol,zic,wdb,bek,nanopo,ram1,roskane,roskane1,cawa,copw}.

In this paper we extend our previous statistical ana\-ly\-sis \cite{bek}
of low energy data to the large \tb region, in which case the bottom
Yukawa couplings cannot be neglected. In addition to the constraints
from gauge and Yukawa coupling unification, electroweak symmetry
breaking and LEP limits on the SUSY mass spectrum, we include now
constraints from $b\to s\gamma$ observed by CLEO~\cite{cleo94} and
the lower limit on the lifetime of the universe, which requires the
dark matter density from the Lightest Supersymmetric Particle (LSP)
not to overclose the universe.

The theoretically more questionable constraint from proton decay in
the MSSM \cite{arn,langac}, which involves the unknown Higgs sector at
the GUT scale will not be considered.  At $\tb < 10$ this constraint
can be fulfilled \cite{bek}, but at large $\tb$ values one needs an
extension of the minimal model, i.e either a different multiplet
structure\cite{flipsu5} or a larger Higgs sector\cite{finite}.
 
Assuming soft symmetry breaking at the \rG scale, all SUSY masses can
be expressed in terms of 5 pa\-ra\-me\-ters and the masses at low energy are
then determined by the well known Renormalization Group Equations
(RGE). The experimental constraints are sufficient to determine these
parameters, albeit with large uncertainties.  The statistical
analysis yields the pro\-ba\-bi\-li\-ty for every point in the SUSY
parameter space, which allows us to calculate the cross sections for
the expected new physics of the MSSM at LEP II.  These cross sections
will be given as function of the common scalar and gaugino masses at
the \rG scale, denoted by $\mze,$ $\mha;$ for each choice of $\mze,$
$\mha,$ the other parameters were determined from the constrained fit.

\section{The Model}
\subsection{The Lagrangian}

The MSSM is completely specified by the standard $SU(3)_C\times
SU(2)_L\times U(1)_Y$ gauge couplings as well as by the low-energy
superpotential and "soft" SUSY breaking terms \cite{rev}.  The most
general gauge invariant form of the R-parity $(R_p = (-1)^{3B+L+2S})$
conserving superpotential is
\begin{equation}
\label{superpot}
{\cal W} = h_{E}L^{j}E^{c}H_{1}^{i}{\epsilon_{ij}}
         + h_{D}Q^{j}D^{c}H_{1}^{i}{\epsilon_{ij}}
         + h_{U}Q^{j}U^{c}H_{2}^{i}{\epsilon_{ij}}
         + \mu H_{1}^{i}H_{2}^{j}{\epsilon_{ij}}
\end{equation}
($\epsilon_{12}=+1$). The following notations are used for the quark
$Q(3,2,1/6)$, $D^{c}(\overline 3,1,1/3)$, $U^{c}(\overline 3,1,-2/3)$,
lepton
$L(1,2,-1/2)$, $E^{c}(1,1,1)$ and Higgs $H_{1}(1,2,-1/2)$,
$H_{2}(1,2,1/2)$ chiral superfields with the $SU(3)_C\times
SU(2)_L\times U(1)_Y$ assignment given in brackets;
$\tilde{m}_U,\tilde{m}_D$ and $\tilde{m}_E$ refer to the masses of the
superpartners of the quark and lepton singlets, while $\tilde{m}_Q$
and $\tilde{m}_L$ refer to the masses of the weak isospin doublet
superpartners. Yukawa coupling constants $h_{E, D, U}$ are
non-diagonal matrices in generation space.  Since the masses of the
third generation are much larger than masses of the first two ones, we
consider only the Yukawa coupling of the third generation and do not
write the generation indices.  In this case $h_{E, D, U}\rightarrow
h_{\tau, b, t}$.

The ``soft'' SUSY breaking terms, by construction, do not generate
quadratic divergences. These terms might originate from supergravity.
In general, the ``soft'' SUSY breaking terms are given by \cite{soft}:

\begin{eqnarray}
\label{soft}
{\cal L}_{\rm SB} &=& - \frac{1}{2} \sum_A M_A
\bar\lambda_A \lambda_A 
- m_{H_1}^2 |H_1|^2 - m_{H_2}^2 |H_2|^2 \\
&& - \tilde m^{2}_Q |\tilde Q|^2 
- \tilde m^2_D |\tilde D^c|^2 - \tilde m^2_U
|\tilde U^c|^2 - \tilde m^2_L
|\tilde L|^2 - \tilde m^2_E |\tilde E^c|^2 \nn 
&& - m_0(h_{E} A_{E} \tilde{L}^{j} \tilde{E}^{c} H_{1}^{i} \epsilon_{ij}
+h_{D} A_{D} {\tilde{Q}}^{j} {\tilde{D}}^{c} H_{1}^{i}\epsilon_{ij} 
 + h_{U} A_{U} {\tilde{Q}}^{j} {\tilde{U}}^{c} H_{2}^{i}
\epsilon_{ij}+h.c) \nn
&&-m_0(B \mu H_{1}^{i} H_{2}^{j} \epsilon_{ij} +h.c)\nonumber
\end{eqnarray}
$A$ is summed over all gauginos: $M_{3,2,1}$ are the masses of the
$SU(3)_C\times SU(2)_L\times U(1)_Y$ gauginos
$\tilde g, \tilde W, \tilde B$ and $m_i$ are the masses
of the scalar fields. $A_{\tau}, A_b, A_t$ and $B$
are trilinear and bilinear coup\-lings.

Assuming Grand Unification results in the following free MSSM
parameters:
\begin{itemize}
\item The common gauge coupling $\alpha_{GUT}$.
\item The matrices of the Yukawa couplings $h_i^{ab}$, where $i = \tau,
  t, b$. 
\item The Higgs field mixing parameter  $\mu $.
\item The soft supersymmetry breaking parameters $m_0$, $m_{1/2}$,
  $A_0$, and $B$.
\end{itemize}
Additional constraints follow from the minimization conditions of the
scalar Higgs potential, which will be discussed later.  Using these
conditions the bilinear coupling $B$ can be replaced by the ratio
$\tb=v_2/v_1$ of the vacuum expectation values of the two Higgs
doublets at the scale $M_Z$.

\subsection{The SUSY Mass Spectrum \label{rge}}

All couplings and masses become scale ($Q^2$) dependent due to
radiative corrections. This running is described by the
renormalization group equations (RGE) \cite{bbo}. 

The following definitions are used: \mgut is the GUT scale, $Q$ is the
running scale, $\alpha_i,~(i=1,2,3)$ are the three gauge
couplings, $\tilde\alpha_i \equiv \alpha_i/4\pi$, $Y_j,~(j=t,b,\tau) 
\equiv h^2_j/(4\pi)^2$ are the third generation Yukawa
couplings, and $t=\ln(\frac{\mgut^2}{Q^2})$.  The couplings $h_j$ are
related to the masses by
\begin{eqnarray}
\label{yukawastbtau} m_{t}&=&h_t(m_t)v\sin\beta \nn m_b & =&
h_b(m_b)v\cos\beta \nn m_\tau&=& h_\tau(m_\tau)v\cos\beta
\end{eqnarray}
Here $m_j$ ($j=t,b,\tau$) are the running masses. 

Defining a vector
$\tilde\alpha_i = \tilde\alpha_1,...,\tilde\alpha_6 =
(\tilde\alpha_1,\tilde\alpha_2,\tilde\alpha_3,Y_t,Y_b,Y_\tau)$
allows to write the RGE in a compact form: \\
{\bf 3 Gauge couplings: ($i=1,2,3$)} \\
\begin{eqnarray}
   \frac{d\tal_{i}}{dt} & = &
            -b_i \tal_i^2 -\tal_i^2\left( \sum_{j=1}^3
            b_{ij}\tal_j- \sum_{j=t,b,\tau}a_{ij} Y_j\right) 
                                             \label{coupl}
\end{eqnarray}
{\bf 3 Yukawa couplings:} \\
\begin{eqnarray}
   \frac{dY_t}{dt}    & = & Y_t
           \sum_{i=1}^6 \left( c^t_i \tal_i -
           \sum_{j\geq i}^6 c^t_{ij} \tal_i \tal_j \right)
                                                \label{ytop} \\
   \frac{dY_b}{dt}    & = & Y_b
           \sum_{i=1}^6 \left( c^b_i \tal_i -
           \sum_{j\geq i}^6 c^b_{ij} \tal_i \tal_j \right)
                                                \label{ybot} \\
   \frac{dY_\tau}{dt} & = & Y_\tau
           \sum_{i=1}^6 \left( c^\tau_i \tal_i -
           \sum_{j\geq i}^6 c^\tau_{ij} \tal_i \tal_j \right) 
\end{eqnarray}
{\bf 3 Gauginos: ($i=1,2,3$)} \\
\begin{eqnarray}
    \frac{dM_{i}}{dt} & = & -b_i\tilde{\alpha}_i^2M_i        \label{RGEM}
\end{eqnarray}
The various coefficients have been summarized in tables
\ref{tbi}--\ref{tctauij}.\\
{\bf Masses of the 1st. and 2nd Generation ($i=1,2$):} \\
\begin{eqnarray}
\frac{d\tilde{m}^2_{L_i}}{dt} & = & 
    3(\tilde{\alpha}_2M^2_2 + \frac{1}{5}\tilde{\alpha}_1M^2_1) \\
\frac{d\tilde{m}^2_{E_i}}{dt} & = & (
     \frac{12}{5}\tilde{\alpha}_1M^2_1) \\
\frac{d\tilde{m}^2_{Q_i}}{dt} & = & (\frac{16}{3}\tilde{\alpha}_3M^2_3
    + 3\tilde{\alpha}_2M^2_2 + \frac{1}{15}\tilde{\alpha}_1M^2_1)\\
\frac{d\tilde{m}^2_{U_i}}{dt} & = & (\frac{16}{3}\tilde{\alpha}_3M^2_3
    +\frac{16}{15}\tilde{\alpha}_1M^2_1)  \\
\frac{d\tilde{m}^2_{D_i}}{dt} & = &
    (\frac{16}{3}\tilde{\alpha}_3M^2_3+
    \frac{4}{15}\tilde{\alpha}_1M^2_1) 
\end{eqnarray}
{\bf Masses of the 3th Generation:} \\ 
\begin{eqnarray}
\frac{d\tilde{m}^2_{L_3}}{dt} & = & 
    3(\tilde{\alpha}_2M^2_2 + \frac{1}{5}\tilde{\alpha}_1M^2_1)
   -Y_\tau(\tilde{m}^2_{L_3}+\tilde{m}^2_{E_3}+m^2_{H_1}+A^2_\tau m_0^2) 
           \label{RGEmL}\\
\frac{d\tilde{m}^2_{E_3}}{dt} & = & (
     \frac{12}{5}\tilde{\alpha}_1M^2_1) -2Y_\tau(\tilde{m}^2_{L_3}+\tilde{m}^2_{E_3}
    +m^2_{H_1}+A^2_\tau m_0^2) \label{RGEmE}\\
\frac{d\tilde{m}^2_{Q_3}}{dt} & = & (\frac{16}{3}\tilde{\alpha}_3M^2_3
    + 3\tilde{\alpha}_2M^2_2 + \frac{1}{15}\tilde{\alpha}_1M^2_1)
      \label{RGEmQ} \\
 && - [Y_t(\tilde{m}^2_{Q_3}+\tilde{m}^2_{U_3}+m^2_{H_2}+A^2_t m_0^2)
     +Y_b(\tilde{m}^2_{Q_3}+\tilde{m}^2_{D_3}+m^2_{H_1}+A^2_b m_0^2)],\nn
\frac{d\tilde{m}^2_{U_3}}{dt} & = & (\frac{16}{3}\tilde{\alpha}_3M^2_3
    +\frac{16}{15}\tilde{\alpha}_1M^2_1)
    -2Y_t(\tilde{m}^2_{Q_3}+\tilde{m}^2_{U_3}+
    m^2_{H_2}+A^2_t m_0^2), \label{RGEmU}\\
\frac{d\tilde{m}^2_{D_3}}{dt} & = &
    (\frac{16}{3}\tilde{\alpha}_3M^2_3+
    \frac{4}{15}\tilde{\alpha}_1M^2_1)
    -2Y_b(\tilde{m}^2_{Q_3}+\tilde{m}^2_{D_3}+m^2_{H_1}+A^2_b m_0^2) 
                  \label{RGEmD}
\end{eqnarray}
{\bf Higgs potential parameters:}\\
\begin{eqnarray}
  \frac{d\mu^2}{dt}&=&\mu^2\left[3(\tilde{\alpha}_2+
    \frac{1}{5}\tilde{\alpha}_1)-(3Y_t+3Y_b+Y_\tau)\right] \\
  \frac{dm^2_{H_1}}{dt} & = &
  3(\tilde{\alpha}_2M^2_2 +\frac{1}{5}\tilde{\alpha}_1M^2_1)
    - 3Y_b(\tilde{m}^2_{Q_3}+\tilde{m}^2_{D_3}+m^2_{H_1} +A^2_b m_0^2) \\
  && - Y_\tau(\tilde{m}^2_{L_3}+\tilde{m}^2_{E_3}+m^2_{H_1} +A^2_\tau m_0^2) \nn
\frac{dm^2_{H_2}}{dt} & = &
3(\tilde{\alpha}_2M^2_2 +\frac{1}{5}\tilde{\alpha}_1M^2_1)
-3Y_t(\tilde{m}^2_{Q_3}+\tilde{m}^2_{U_3}+m^2_{H_2}+A^2_t m_0^2)
\end{eqnarray}
{\bf Trilinear couplings:} \\
\begin{eqnarray}
\frac{dA_t}{dt} & = &
  -\left(\frac{16}{3}\tilde{\alpha}_3\frac{M_3}{m_0} +
  3\tilde{\alpha}_2\frac{M_2}{m_0} +
  \frac{13}{15}\tilde{\alpha}_1\frac{M_1}{m_0}\right) -6Y_tA_t-Y_bA_b\\
\frac{dA_b}{dt} & = &
  -\left(\frac{16}{3}\tilde{\alpha}_3\frac{M_3}{m_0}
  + 3\tilde{\alpha}_2\frac{M_2}{m_0} +
  \frac{7}{15}\tilde{\alpha}_1\frac{M_1}{m_0}\right) -6Y_bA_b-Y_tA_t-Y_\tau A_\tau\\
\frac{dA_\tau}{dt} & = & -\left(
 3\tilde{\alpha}_2\frac{M_2}{m_0} +
\frac{9}{5}\tilde{\alpha}_1\frac{M_1}{m_0}\right) -3Y_bA_b-4Y_\tau A_\tau
%
%
\end{eqnarray}

Here $A_t,A_b,A_\tau$ and $B$ are the couplings in ${\cal L}_{\rm SB}$ 
as defined before; $M_i$ are the gaugino masses before
any mixing. The boundary conditions at $Q^2=M_{\rm
  GUT}^2$ or at $t=0$ are:
$$m^2_{H_1}=m^2_{H_2}=\tilde{m}^2_Q=\tilde{m}^2_U=\tilde{m}^2_D=
  \tilde{m}^2_L=\tilde{m}^2_E= m_0^2;$$
$$\mu^2 = \mu_0^2;\quad M_i=m_{1/2};
\quad \tal_i(0)=\tal_{\rm GUT},\quad i=1,2,3$$
$$A_t=A_b=A_\tau=A_0.$$
With given
values for $m_0,m_{1/2},\mu,Y_t,Y_b,Y_\tau,\tan\beta$, and $A_0$ and
correspondingly known boundary conditions at the GUT scale, the
differential equations can be solved numerically, thus linking the
values at the GUT and electroweak scales. The non-negligible Yukawa
couplings cause a mixing between the electroweak eigenstates and the
mass eigenstates of the third generation particles.  The mixing
matrices are:
\begin{equation} \label{stopmat}
\left(\begin{array}{cc}
\tilde{m}^2_{Q_3}+m_t^2+\frac{1}{6}(4M_W^2-M_Z^2)\cos 2\beta & 
m_t(A_tm_0-\mu\cot \beta ) \\  
m_t(A_tm_0-\mu\cot \beta ) &   
\tilde{m}^2_{U_3}+m_t^2-\frac{2}{3}(M_W^2-M_Z^2)\cos 2\beta
\end{array}  \right)       \nonumber
\end{equation}
\begin{equation} \label{botmat}
\left(\begin{array}{cc}
\tilde{m}_{Q_3}^2+m_b^2-\frac{1}{6}(2M_W^2+M_Z^2)\cos 2\beta & 
m_b(A_bm_0-\mu\tan \beta ) \\  
m_b(A_bm_0-\mu\tan \beta ) &   
\tilde{m}_{D_3}^2+m_b^2+\frac{1}{3}(M_W^2-M_Z^2)\cos 2\beta 
\end{array}  \right)               \nonumber
\end{equation}
\begin{equation} \label{staumat}
\left(\begin{array}{cc}
\tilde{m}_{L_3}^2+m_{\tau}^2-\frac{1}{2}(2M_W^2-M_Z^2)\cos 2\beta & 
m_{\tau}(A_{\tau}m_0-\mu\tan \beta ) \\  
m_{\tau}(A_{\tau}m_0-\mu\tan \beta ) &   
\tilde{m}_{E_3}^2+m_{\tau}^2+(M_W^2-M_Z^2)\cos 2\beta
\end{array}  \right)               \nonumber
\end{equation}
and the  mass eigenstates are  
 the eigenvalues of these mass matrices. 
The mass matrix for the  neutralinos can be written in our notation
as:
\begin{equation} 
{\cal M}^{0}= 
\left(
  \begin{array}{cccc}
    M_1        & 0           & -M_Z \cos\beta \sin\theta_W  & M_Z \sin\beta \sin\theta_W  \\
    0          & M_2         &  M_Z \cos\beta \cos\theta_W &-M_Z \sin\beta \cos\theta_W \\
   -M_Z \cos\beta \sin\theta_W &  M_Z \cos\beta \cos\theta_W & 0           & -\mu \\
    M_Z \sin\beta \sin\theta_W & -M_Z \sin\beta \cos\theta_W &-\mu         & 0
  \end{array} \right) \label{neutmix} 
\end{equation}

The physical neutralino masses  $\tilde{m}_{\chi_i^0}$
are obtained as eigenvalues of this matrix after diagonalization.
The mass matrix for the charginos is:
\begin{equation}
  {\cal M}^{\pm}=\left(
  \begin{array}{cc}
     M_2                  & \sqrt{2}M_W\sin\beta \\
     \sqrt{2}M_W\cos\beta & \mu
  \end{array} \right) \label{charmix}
\end{equation}
This matrix has two eigenvalues corresponding to the masses of the
two charginos~ $\tilde{\chi}_{1,2}^{\pm}$:
\begin{eqnarray}
 \tilde{m}^2_{\chi^\pm_{1,2}}&=&\frac{1}{2}\bigl[M^2_2+\mu^2+2M^2_W \\
 && \mp \sqrt{(M^2_2-\mu^2)^2+4M^4_W\cos^22\beta 
 +4M^2_W(M^2_2+\mu^2+2M_2\mu \sin 2\beta)}\bigr]\nonumber
\end{eqnarray}

\subsection{Radiative Corrections to the Higgs potential} 
\label{cor_higgs_pot}
The Higgs potential $V$ for the neutral components including the
one-loop corrections $\Delta V$ can be written
as\cite{loopewbr}:
\begin{eqnarray}
  V(H_1^0,H_2^0) &=& m^2_1|H_1^0|^2+m^2_2|H_2^0|^2 -m^2_3(H_1^0H_2^0+h.c.)\\
                 & & +\frac{g^2+g^{'2}}{8}(|H_1^0|^2-|H_2^0|^2)^2 + \Delta V\nn
{\rm with }\quad\Delta V&=&\frac{1}{64\pi^2}
\sum_i(-1)^{2J_i}(2J_i+1) C_im_i^4\left[\ln\frac{m_i^2}{Q^2}-\frac{3}{2}\right],\label{21} \end{eqnarray}
where the mass parameters are defined as
\begin{eqnarray}
m^2_1&=&m^2_{H_1}+\mu^2\\ 
m^2_2&=&m^2_{H_2}+\mu^2\\
m^2_3&=&B m_0 \mu
\end{eqnarray}
and the sum is taken over all possible particles (masses $m_i$) inside
the loops.

The minimization conditions
$$\frac{\partial V}{\partial\psi_1}=0, \quad
\frac{\partial V}{\partial\psi_2}=0$$
with $\psi_{1,2}={\bf Re}H^0_{1,2}$ yield:
\begin{eqnarray}
2m_1^2&=&2m_3^2\tan \beta - M_Z^2\cos 2\beta - 2\Sigma_1 \label{2m1} \\
2m_2^2&=&2m_3^2\cot \beta + M_Z^2\cos 2\beta - 2\Sigma_2,\label{2m2} 
\end{eqnarray} 
where $\Sigma_1\equiv \frac{1}{2}\frac{\partial \Delta
  V}{\partial\psi_1} $ and    
$\Sigma_2\equiv \frac{1}{2}\frac{\partial \Delta V}{\partial\psi_2} $ 
are the one-loop corrections\cite{loopewbr}:
\begin{eqnarray} 
\Sigma_1   & = & \frac{1}{64\pi^2}
\sum_i (-1)^{2J_i}(2J_i+1)\frac{1}{\psi_1}\frac{\partial m_i^2}
{\partial \psi_1}f(m_i^2) \label{sigma1} \\
\Sigma_2 & = & \frac{1}{64\pi^2} 
\sum_i (-1)^{2J_i}(2J_i+1)\frac{1}{\psi_2}\frac{\partial m_i^2}
{\partial \psi_2}f(m_i^2) \label{sigma2}
\end{eqnarray}
and the function $f$ is defined as\footnote{This definition differs by
  a factor 2 from the one of Ellis et al.~\cite{erz}}:
\begin{eqnarray} 
f(m^2) & = & m^2\left(\ln\frac{m^2}{Q^2}-1\right) 
\end{eqnarray}
The Higgs masses can now be calculated including the complete 1-loop
corrections for given masses $m_i$ \cite{cpr,bekhiggs}. The dominant
2-loop corrections from the third generation have been calculated in
refs. \cite{erz,berz,kz92,eqz}.
%
%
%
%
\section{Comparison of the MSSM with experimental Data}
 In this section the various low energy GUT predictions are
compared with data. The most  restrictive constraints are
the coupling constant unification and the requirement that the
unification scale has to be above $10^{15}$ GeV from the proton
lifetime limits, assuming decay via s-channel exchange of heavy
gauge bosons. They exclude the SM~\cite{ekn2,abf,lalu} as well
as many other models~\cite{abf,abfI,yana}. The only model known to
be able to fulfill all constraints simultaneously is the MSSM.
In the following we shortly summarize the experimental inputs
and then discuss the fit results.

\subsection{Coupling Constant Unification}
The three coupling constants of the known symmetry groups are: 
\begin{equation}
\label{SMcoup}{\matrix{
\alpha_1&=&(5/3)g^{\prime2}/(4\pi)&=&5\alpha/(3\cos^2\theta_W)\cr
\alpha_2&=&\hfill g^2/(4\pi)&=&\alpha/\sin^2\theta_W\hfill\cr
\alpha_3&=&\hfill g_s^2/(4\pi)\cr}}
\end{equation}
where $g',g$ and $g_s$ are the $U(1)$, $SU(2)$ and $SU(3)$ coupling 
constants.

The couplings, when defined as effective values including loop
corrections in the gauge boson propagators, become energy dependent
(``running'').  A running coupling requires the specification of a
renormalization prescription, for which the modified minimal
subtraction ($\overline{MS}$) scheme~\cite{msbar} is used.

In this scheme the world averaged values of the coup\-lings at the
Z$^0$ energy are obtained from a fit to the LEP data~\cite{LEP}, $M_W$
\cite{PDB} and \mt\cite{CDF,D0}:
\begin{eqnarray}
  \label{worave}
  \alpha^{-1}(M_Z)             & = & 128.0\pm0.1\\
  \sin^2\theta_{\overline{MS}} & = & 0.2319\pm0.0004\\
  \alpha_3                     & = & 0.125\pm0.005.
\end{eqnarray}
The value of $\alpha^{-1}(M_Z)$ was updated from ref. \cite{dfs} by
\mbox{using} new data on the hadronic vacuum polarization\cite{EJ95}.
For SUSY models, the dimensional reduction $\overline{DR}$ scheme is a
more appropriate renormalization scheme~\cite{akt}.  In this scheme
all thresholds are treated by simple step approximations and
unification occurs if all three $\aii(\mu)$ meet exactly at one point.
This crossing point corresponds to the mass of the heavy gauge bosons.
The $\overline{MS}$ and $\overline{DR}$ couplings differ by a small
offset
\begin{equation}
\label{MSDR}{{1\over\alpha_i^{\overline{DR}}}=
{1\over\alpha_i^{\overline{MS}}}-{C_i\over\strut12\pi}},
\end{equation}
where the $C_i$ are the quadratic Casimir coefficients of the group
($C_i=N$ for SU($N$) and 0 for U(1) so $\alpha_1$ stays the same).  In
the following the $\overline{DR}$ scheme will be used.
\subsection{\mz from Electroweak Symmetry Breaking}
Radiative corrections trigger spontaneous symmetry breaking in the
electroweak sector.  In this case the Higgs potential does not have
its minimum for all fields equal zero, but the minimum is obtained for
non-zero vacuum expectation values of the fields.  Solving $\mz$ from
eqns. \ref{2m1} and \ref{2m2} yields: 
\begin{equation}
\label{defmz}
\frac{\mz^2}{2}=\frac{m_1^2+\Sigma _1-(m_2^2+\Sigma _2) \tan^2\beta}
{\tan^2\beta-1},
\end{equation}
where the $\Sigma_1$ and $\Sigma_2$ are defined in
eqns.~\ref{sigma1} and \ref{sigma2}.  
\subsection{Yukawa Coupling~Constant Unification}
\label{sec:masses}
The masses of top, bottom and $\tau$ can be obtained from the low
energy values of the running Yukawa couplings as shown in
eq.~(\ref{yukawastbtau}). The requirement of $b-\tau$ Yukawa coupling
unification strongly restricts the possible solutions in the $\mt$
versus $\tb$ plane, as discussed by many groups
\cite{Ross1,bbog,lanpol,bmaskln,cpr,copw1,ara91}.  The values of the
running masses can be translated to pole masses following the formulae
from~\cite{runmas}.  In the MSSM the bottom mass has additional
corrections from loops involving squark gluino and stop chargino
loops~\cite{hemp,hall}.  These corrections are small for low \tb
solutions, but become large for the high \tb values.

\begin{eqnarray}
\label{deltab}
      \Delta m_b  & = & \frac{2\alpha_3}{3\pi}\;m_{\tilde g}\;\mu\;\tb\;
{\rm I}(\smas{b_1}{2},\smas{b_2}{2},\smas{g}{2})
          + Y_t \; A_t   \; \mu \; \tb \;
          {\rm I}(\smas{t_1}{2},\smas{t_2}{2},\mu),\\
{\rm I}(x,y,z) &=& \frac{-[xy\log(x/y)+yz\log(y/z)+zx\log(z/x)]}
                        {(x-y)(y-z)(z-x)} 
\end{eqnarray}
This corrections, proportional to $\tan\beta$, are added to $m_b$:
\begin{eqnarray}
m_b^{MSSM}(M_Z^2) & = & m_b(M_Z^2)(1+\Delta m_b)
\end{eqnarray}
Corresponding corrections also exist for the $\tau$ lepton:
\begin{eqnarray}
   \Delta m_\tau & = & \frac{\alpha_1}{4\pi}\;m_{\tilde B}\;\mu\;\tb \;
        {\rm I}(\smas{\tau_1}{2},\smas{\tau_1}{2},\smas{B}{2}),
\end{eqnarray}
but they are negligible to $\Delta m_b$, because they are
proportional to $\alpha_1$ and the bino mass $\tilde{m}_B$,
which gives a suppression of a factor of $\approx 0.01$.

For the pole masses of the third generation the following values   
are taken:
\begin{eqnarray}
 M_t    & = & 179     \pm 12~     \mbox{GeV}/c^2~\mbox{\cite{CDF,D0}},     \nn
 M_b    & = &   4.94  \pm  0.15~  \mbox{GeV}/c^2~\mbox{\cite{bmass}},      \nn
 M_\tau & = &   1.7771\pm  0.0005~\mbox{GeV}/c^2~\mbox{\cite{taumass,PDB}}.
\end{eqnarray}
Since the gauge couplings are measured most precisely at $\mz$, the
Yukawa couplings were fitted at $\mz$ too.  The running mass of the
$b$-quark at $\mz$ was calculated by using the third order QCD
formula\cite{ckw93}, which leads to $m_b(\mz)=2.84\pm0.15~GeV/c^2$ for
$\as(\mz)=0.125\pm0.005$; the error on $m_b$ includes the uncertainty
from $\as$.  The running of $m_\tau$ is much less between $M_\tau$ and
$\mz$; one finds $m_\tau(\mz) = 1.7462\pm0.0005$. The Yukawa coupling
of the top quark is always evaluated at $M_t$, since its running
depends on the SUSY spectrum, which may be splitted in particles below
and above $M_t$.

\subsection{Branching Ratio $BR(b\to s \gamma)$}\label{sec:bsg}
The branching ratio \Bbsg has been measured by the CLEO collaboration
\cite{cleo94} to be: $BR(b\to s \gamma)=2.32\pm0.67\times10^{-4}$.

In the MSSM this flavour changing neutral current (FCNC) receives in
addition to the SM $W-t$ loop contributions from $H^\pm-t$,
$\tilde{\chi}^\pm - \tilde{t}$ and $\tilde{g}-\tilde{q}$ loops.  The
$\tilde{\chi}^0 -\tilde{t}$ loops, which are expected to be much
smaller, have been neglected\cite{borz,bsgamm3}. The
$\tilde{g}-\tilde{q}$ loops are proportional to \tb.  From the
formulae given by Oshima \cite{bsgamm3} we found this contribution to
be small, even in the case of large \tb and therefore it was neglected.
The chargino contribution, which becomes large for large $\tb$ and
small chargino masses, depends sensitively on the splitting of the two
stop masses; therefore it is important to diagonalize the matrix
without approximations.

The theoretical prediction depends on the renor\-ma\-li\-za\-tion
scale~\cite{burasb} for the standard QCD corrections to this decay.
Varying this scale between $m_b/2$ and $2m_b$ leads to a theoretical
uncertainty $\sigma_{th.}=0.6\times10^{-4}$, which is added in
quadrature to the experimental error. The fit prefers scales close to
the upper limit, so the analysis was done with $2m_b$ as
renormalization scale.

Within the MSSM the following ratio has been calculated
\cite{bsgamma,bsgamm3}:
\begin{equation}
\frac{BR(b\to s\gamma)}{BR(b\to c e
\bar{\nu})}=\frac{|V_{ts}^*V_{tb}|^2}{|V_{cb}|^2}
K_{NLO}^{QCD}
\frac{6\alpha}{\pi}
\frac{\left[\eta^{16/23}A_\gamma+\frac{8}{3}(\eta^{14/23}
-\eta^{16/23})A_g+C\right]^2}{I(m_c/m_b)
[1-(2/3\pi)\alpha_s(m_b)f(m_c/m_b)]},\nonumber
\end{equation} where
\begin{eqnarray}
C &\approx& 0.175, \ \ I = 0.4847, 
\\ \eta&=&\alpha_s(M_W)/\alpha_s(m_b), \ \
 f(m_c/m_b)=2.41. 
\end{eqnarray}
Here $ f(m_c/m_b)$ represents corrections from {\em leading order} QCD
to the known semileptonic $b\to c e \bar{\nu}$ decay rate, while the
ratio of masses of $c$- and $b$-quarks is taken to be $m_c/m_b=0.316$. The
ratio of CKM matrix elements
$\frac{|V_{ts}^*V_{tb}|^2}{|V_{cb}|^2}=0.95$ was taken from Buras et
al. \cite{burasb}  the {\em next leading order } QCD-Corrections from Ali et
al.~\cite{alibsg}. $A_{\gamma,g}$ are the coefficients of the
effective operators for $bs$-$\gamma$ and for $bs$-gluon interactions
respectively. Using the formulae of ref.~\cite{bsgamma} to compare with
the experimental results leads to significant constraints on the
parameter space, especially at large values of $\tb$, as discussed in
refs. \cite{bop,cawa,roskane,bsgamm1,bsgamm2}.

For large $\tb$ the chargino contribution is dominant 
and is proportional to $A_t\mu$ \cite{cawa}
\begin{eqnarray}
\label{chargino}
  A_{\gamma,g} & \sim &
       \frac{m^2_t}{m^2_{\tilde{t}}}
       \frac{A_t\mu}{m^2_{\tilde{t}}}\tan\beta.
\end{eqnarray}
For positive (negative) values of $A_t\mu$ this leads to a larger
(smaller) branching ratio \Bbsg as for the Standard Model with two
Higgs doublets.

\subsection{Experimental Lower Limits on SUSY Masses}
SUSY particles have not been found so far and from the searches at LEP
one knows that the lower limit on the charged leptons and charginos is
about half the $Z^0$ ~mass (45 GeV)~\cite{PDB} and the Higgs mass has
to be above 60 GeV~\cite{higgslim,sopczak}. The lower limit on the
lightest neutralino is 18.4 GeV~\cite{PDB}, while the sneutrinos have
to be above 41 GeV~\cite{PDB}. From the short LEP II run at 130 GeV in
November 1995 the lower limit on the chargino mass is 65 GeV
\cite{lepcol}.  These limits require minimal values for the SUSY mass
parameters.  There exist also li\-mits on squark and gluino masses
from the hadron colliders~\cite{PDB}, but these limits depend on the
assumed decay modes.  Furthermore, if one takes the limits given above
into account, the constraints from the limits on all other particles
are usually fulfilled, so they do not provide additional reductions of
the parameter space in case of the {\it minimal} SUSY model.

\subsection{Dark Matter Constraint }\label{s67}
Abundant evidence for the existence of non-relativistic, neutral,
non-baryonic dark matter exists in our universe\cite{borner,kolb}.
The lightest supersymmetric particle (LSP) is supposedly stable and
would be an ideal candidate for dark matter.

The present lifetime of the universe is at least $10^{10}$ years,
which implies an upper limit on the expansion rate and correspondingly
on the total relic abundance.  Assuming $h_0>0.4$ one finds that the
contribution of each relic particle species $\chi$ has to
obey~\cite{kolb}:
\begin{equation}
\Omega_\chi h^2_0<1,
\end{equation}
where $\Omega_\chi h^2$ is the ratio of the relic particle density of
particle $\chi$ and the critical density, which overcloses the
universe.  This bound can only be met, if most of the LSP's annihilated
into fermion-antifermion pairs, which in turn would annihilate into
photons again.

Since the neutralinos are mixtures of gauginos and higgsinos, the
annihilation can occur both, via s-channel exchange of the $Z^0$ and
Higgs bosons and t-channel exchange of a scalar particle, like a
selectron \cite{relic}.  This constrains the parameter space, as
discussed by many groups\cite{relictst,roskane,rosdm,bop}.  The size
of the Higgsino component depends on the relative sizes of the
elements in the mixing matrix (eq.~\ref{neutmix}), especially on the
mixing angle $\tb$ and the size of the parameter $\mu$ in comparison
to $M_1\approx 0.4m_{1/2}$ and $M_2\approx 0.8 m_{1/2}$.  This mixing
becomes large for the SO(10) type solutions, in which case the
parameters can alway be tuned such, that the relic density is low
enough.

However, for low $\tb$ values the mixing is very small due to the
large value of $\mu$ required from electroweak symmetry breaking and
one finds that the lightest scalars have to be below a few 100 GeV in
that case, as will be discussed below.  The relic density was computed
from the formulae by Drees and Nojiri~\cite{drno} and from the more
approximate formulae by Ellis et al.~\cite{lspdark}.  They typically
agree within a factor two, which is satisfactory and good enough,
since the relic density is such a steep function of the parameters for
low $\tb$, that the excluded regions are hardly changed by a factor
two uncertainty.

\subsection{Fit Method}
\begin{table*}[h]
\begin{center}
\normalsize
\begin{tabular}{|c||c||c|c|}
\hline
                     &&
                       \multicolumn{2}{|c|}{Fit parameters}      \\
\cline{3-4}
\rb{exp.~input data} & \rb{$\Rightarrow$} & low $\tb$
                                               & high $\tb$     \\
\hline
$\alpha_1,\alpha_2,\alpha_3$ &   & \mgut,~\agut  & \mgut,~\agut\\
\mt                          &   & $Y_t^0,~Y_b^0=Y_\tau^0$ &
                                   $Y_t^0=Y_b^0=Y_\tau^0$ \\
\mb              &\rb{ minimize}& \mze,\mha & \mze,\mha  \\
 $m_\tau$        &\rb{ $\chi^2$} & \tb &    \tb      \\
\mz                          &   & $\mu$ &  $\mu$      \\
\bsg                         &   & $(A_0)$ & $A_0$ \\
 $\tau_{universe}$           &   &             &               \\
\hline
\end{tabular} \end{center}
\caption[]{\label{t1}Summary of fit input and output variables. For 
the low $\tb$ scenario the parameter $A_0$ is not very relevant as
indicated by the brackets. For large $\tb$ $\tau_{universe}$ does not
yield any constraints (see text).}
\end{table*}

The fit method has been described in detail before~\cite{bek} for the
low $\tb$ region. In that case the ana\-ly\-ti\-cal solutions for the
SUSY masses could be found and one had to integrate only four RGE's
($\tilde{\alpha}_1$, $\tilde{\alpha}_2$, $\tilde{\alpha}_3$ and $Y_t$)
numerically. For large $\tb$
values all 25 RGE's of section \ref{rge} have to be integrated
simultaneously. As a check, this integration was performed for
low $\tb$ values too and found to be in good agreement with the
results using the analytical solutions for the masses. In the
present analysis the following $\chi^2$ definition is used:
\begin{eqnarray}
 \chi^2 & = &
{\sum_{i=1}^3\frac{(\aii(\mz)-\alpha^{-1}_{MSSM_i}(\mz))^2}
{\sigma_i^2}}
+\frac{(\mz-91.18)^2}{\sigma_Z^2}+{\frac{(M_t -  179)^2}{\sigma_t^2}}\\
& & +\frac{(M_b-4.94)^2}{\sigma_b^2}+\frac{(M_\tau-1.7771)^2}{\sigma_\tau^2} 
+{\frac{(Br(b\to s\gamma)-2.32 \times 10^{-4})^2} {\sigma(b\to s\gamma)^2}} \nn
 & &+{\frac{(\Omega h^2-1)^2}{\sigma^2_\Omega}}\qquad(for ~\Omega h^2 > 1) \nn
 & &+{\frac{(\tilde{M}-\tilde{M}_{exp})^2}{\sigma_{\tilde{M}}^2}}
 \qquad{(for~\tilde{M} < \tilde{M}_{exp})}  \nn
 & &+{\frac{(\tilde{m}_{LSP}-\tilde{m}_{\chi})^2}{\sigma_{LSP}^2}}
 \qquad{(for~ \tilde{m}_{LSP}~charged )}.\nonumber                  \label{chi2}
\end{eqnarray}
The first six terms are used to enforce gauge coupling unification, electroweak
symmetry breaking and $b-\tau$ Yukawa coupling unification, respectively.
The following two terms impose the constraints from \bsg~ and
the relic density, while the 
 last terms require the SUSY masses to be above the 
experimental lower limits and  the lightest supersymmetric 
particle (LSP) to be  a neutralino, since a charged stable LSP would have
been observed. 
The input and fitted output variables have been summarized
in table \ref{t1}.

\section{Results}

\subsection{Constraints from $b-\tau$ unification}
\label{sec:su5}

\begin{figure*}
 \begin{center}
  \leavevmode
  \epsfxsize=15cm
  \epsffile{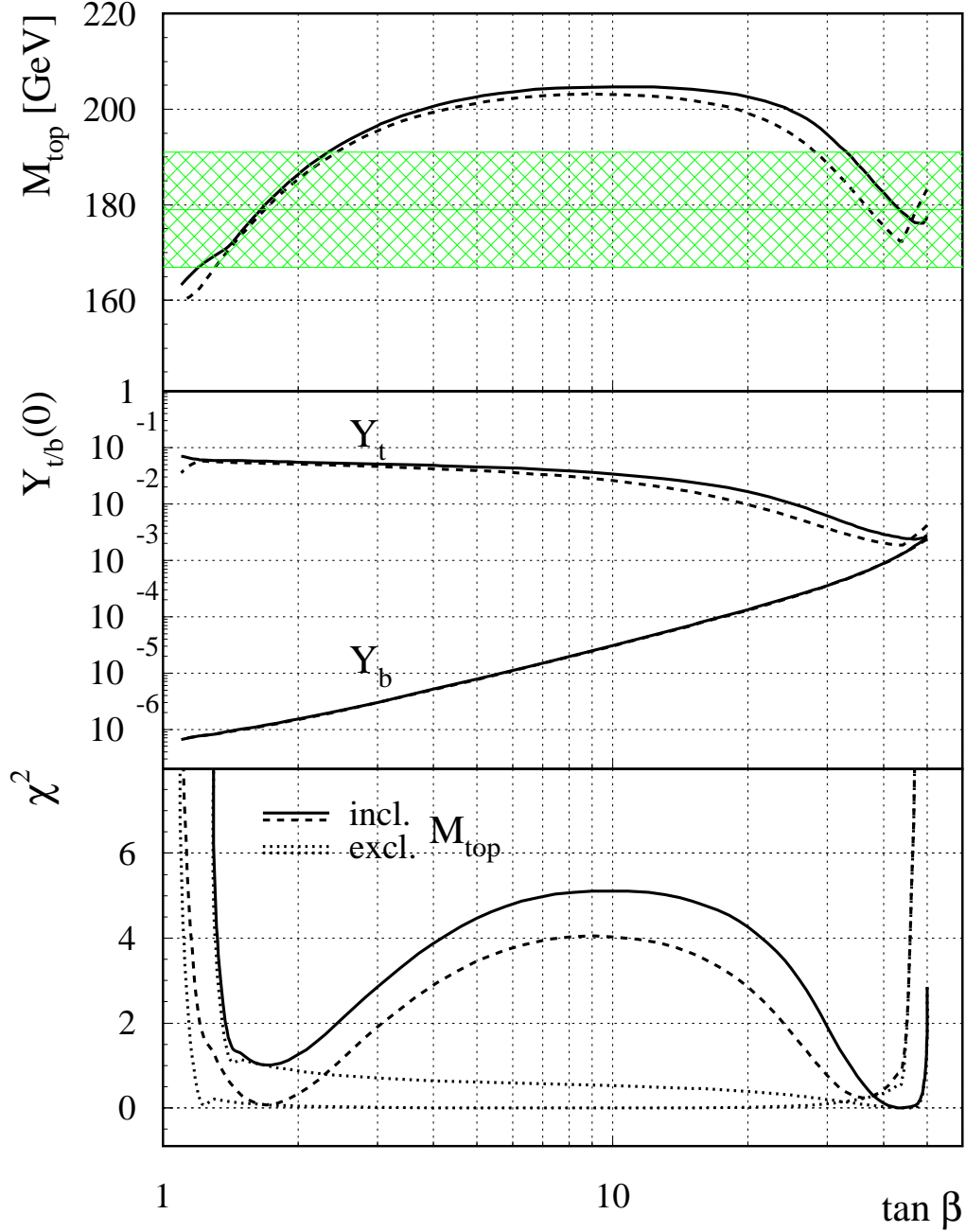}
\end{center}
\caption{\label{\figI}The top quark mass as function of $\tb$ (top) 
for values of $\mze,\mha$ optimized for low and high $\tb$, as indicated
by the dashed and solid lines, respectively.
The middle part shows the corresponding values of the Yukawa
coupling at the GUT scale and the lower part the obtained 
$\chi^2$ values. If the top constraint ($\mt=179\pm12$, horizontal band) 
is not applied, all values of $\tb$ between 1.2 and 50 are allowed
(thin dotted lines at the bottom), but if the top mass 
is constrained to the experimental value, only the regions
$1<\tb<3$ and $25<\tb<50$ are allowed.}
\end{figure*}

The requirement of $b-\tau$ Yukawa coupling unification strongly
restricts the possible solutions in the $M_t$ versus $\tb$ plane, as
discussed before.  With the top mass measured by the CDF and
D0-Collaborations~\cite{CDF,D0} only two regions of $\tb$ give an
acceptable $\chi^2$ fit, as shown in the bottom part of
fig.~\ref{\figI} for two values of the SUSY scales $\mze,\mha$, which
are optimized for the low and high $\tb$ range, respectively, as will
be discussed below.  The curves at the top show the solution for $M_t$
as function of $\tb$ in comparison with the experimental value of
$M_t=179\pm 12$ GeV. The $M_t$ predictions were obtained by imposing
gauge coupling unification and electroweak symmetry breaking for each
value of $\tb$, which allows a determination of $\mu, ~\agut$, and
$\mgut$ from the fit for the given choice of $\mze,\mha$. 
The results do not depend very much on this choice, as can be seen
from a comparison of the solid and dotted lines in fig.~\ref{\figI}.
 
The best $\chi^2$ is obtained for $\tb=1.7$ and $\tb=42$,
respectively.  They correspond to solutions where $Y_t\gg Y_b$ and
$Y_t\approx Y_b$, as shown in the middle part of fig.~\ref{\figI}.
The latter solution is the one typically expected for the SO(10)
symmetry, in which the up and down type quarks as well as leptons
belong to the same multiplet, while the first solution corresponds to
$b-\tau$ unification only, as expected for the minimal SU(5) symmetry.

\subsection{Electroweak symmetry breaking}
\label{sec:elwsb}

Eq.~(\ref{defmz}) for $\mz$ can be written as
\begin{eqnarray} \label{defmz_2}
  \tan^2\beta & = & \frac{m_1^2+\Sigma_1 +
    \frac{1}{2}M_Z^2} {m_2^2 + \Sigma_2 + \frac{1}{2}M_Z^2}.
\end{eqnarray}

\begin{figure*}
 \begin{center}
  \leavevmode
  \epsfxsize=15cm
  \epsffile{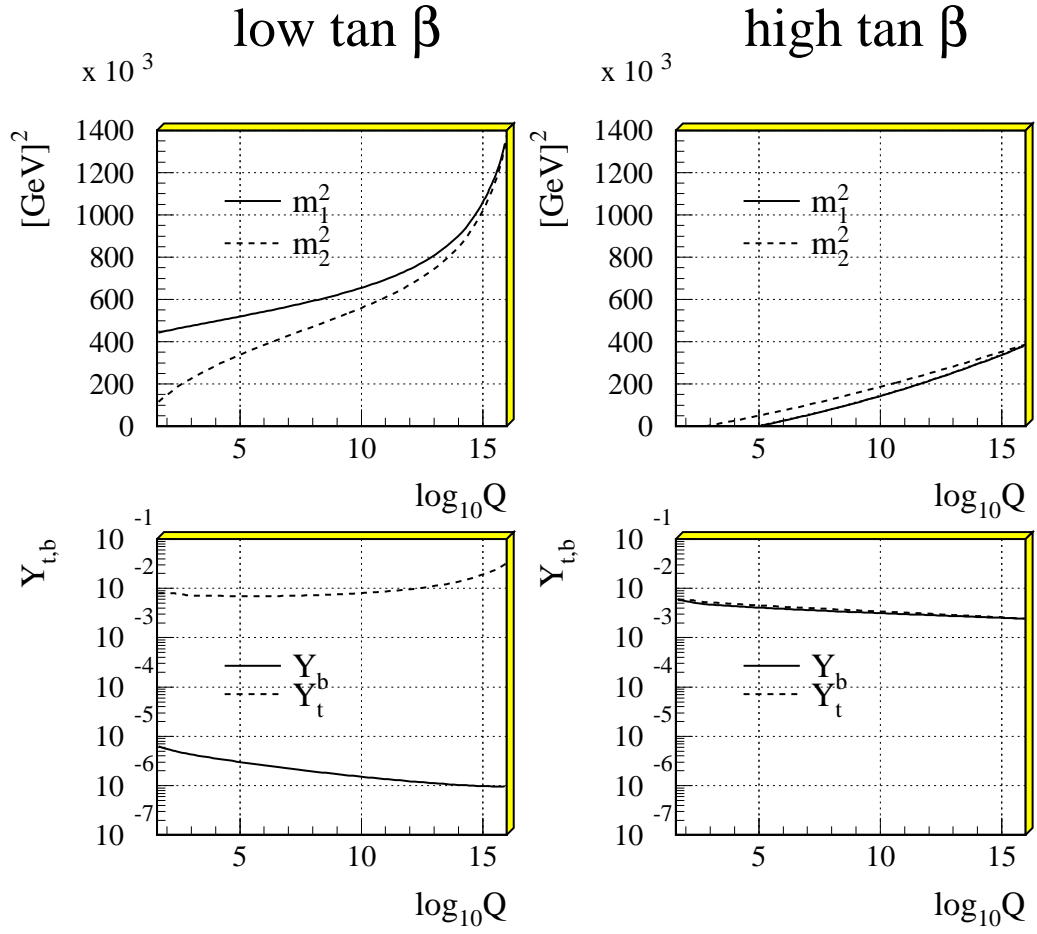}
\end{center}
\caption{\label{\figII}The running of the parameters $m_1$ and $m_2$ in
  the Higgs potential (top) and Yukawa couplings of top and bottom
  quarks (bottom).}
\end{figure*}

From fig.~\ref{\figII} one observes that at low energy $m_1^2 +
\Sigma_1 > m_2^2 + \Sigma_2$  as expected, since $\Sigma_2$
($\Sigma_1$) contains large negative corrections proportional to $Y_t$
($Y_b$) and $Y_t \gg Y_b$. Consequently the numerator in
eq.~\ref{defmz_2} is always larger than the denominator, implying 
$\tan^2\beta > 1$. 

For the large $\tan\beta$ scenario $Y_t \approx Y_b$, so $\Sigma_1
\approx \Sigma_2$ (see fig.~\ref{\figII}). Eq.~\ref{defmz_2} then
requires the starting values of $m_1$ and $m_2$ to be different in
order to obtain a large value of $\tan\beta$. 

Several reasons for such a splitting could be thought of. E.g. SO(10)
may be broken at a scale \msoten above \mgut but below $\mplanck.$ Such
a symmetry breaking can lead to a direct splitting between $m_1$ and
 $m_2$ \cite{carwag94,rattazzi94,rattazzi95}. In addition $Y_t$ and
 $Y_b$ may be different at \mgut due to the evolution between \msoten
and \mgut.

Using the RGE for the SU(5) group from ref. \cite{polon}, we found
$Y_t \approx 1.05 Y_b$ and $m_1 \approx 0.96 m_2$ for the best fitted
value of $\msoten \approx 6\cdot 10^{17}$ GeV. However these
splittings are not large enough to generate electroweak symmetry
breaking.

Good fits can be obtained by splittings between $m_1^2$ and $m_2^2$
and $Y_t$ and $Y_b$ of the order of 25\%, which we used in the
following analysis ($Y_b(0)=0.75 Y_t(0)$, $m_1^2=1.25 m_0^2 + \mu^2$, 
$m_2^2=m_0^2+\mu^2$ at the GUT scale).

The Higgs mixing parameter $\mu$ can be determined from radiative
electroweak symmetry breaking (EWSB), since eq.~\ref{defmz} can be
rewritten as:
\begin{equation}      \label{mufrommz}
 \mu^2 =\frac{m^2_{H_1} - m^2_{H_2}\tan^2\beta}
                  {\tan^2\beta -1}+ \frac{\Sigma_1  - \Sigma_2 \tan^2\beta}
                  {\tan^2\beta -1}- \frac{M_Z^2}{2}\;.
\end{equation}
The dependence of $\mu$ on $m_0$ and $m_{1/2}$ is shown in
fig.~\ref{\figXVI}.  Due to the strong $\tb$ dependence in
eq.~\ref{mufrommz}, the values are much smaller for the high $\tb$
scenario.  One observes that EWSB breaking determines only $\mu^2$, so
the mixing in the stop sector can be either large or small, depending
on the relative sign of $\mu$ and $A_t$.
 
\begin{figure*}
\begin{center}
  \leavevmode
  \epsfxsize=15cm
  \epsffile{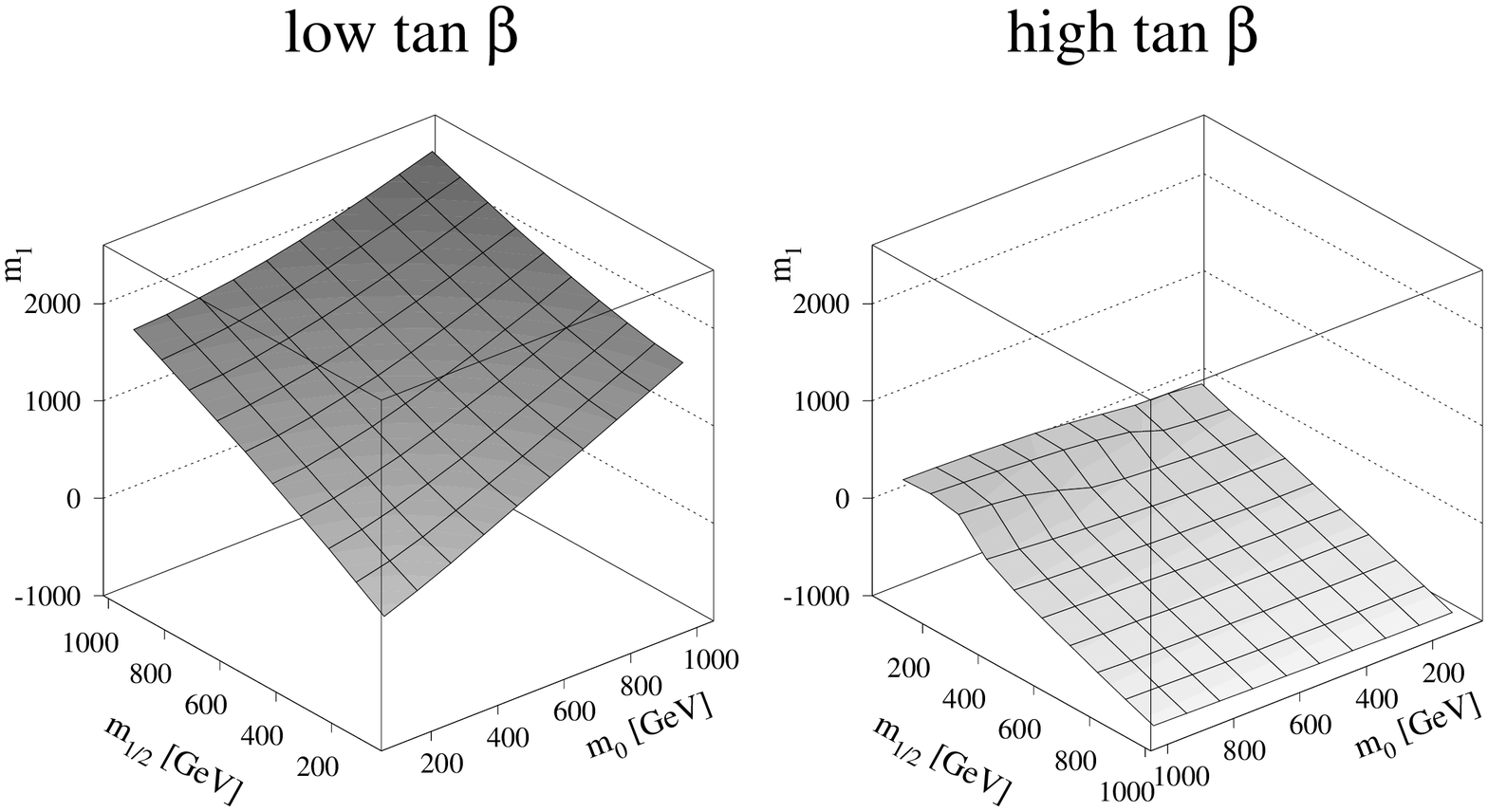}
\end{center}
\caption{\label{\figXIXa}The mass   $m_1$ in the Higgs potential 
  at $\mz$ (Born level) in GeV as function of $\mze$ and $\mha$ for
  the low and high $\tb$ scenario, respectively.  }
\end{figure*}

\begin{figure*}
\begin{center}
  \leavevmode
  \epsfxsize=15cm
  \epsffile{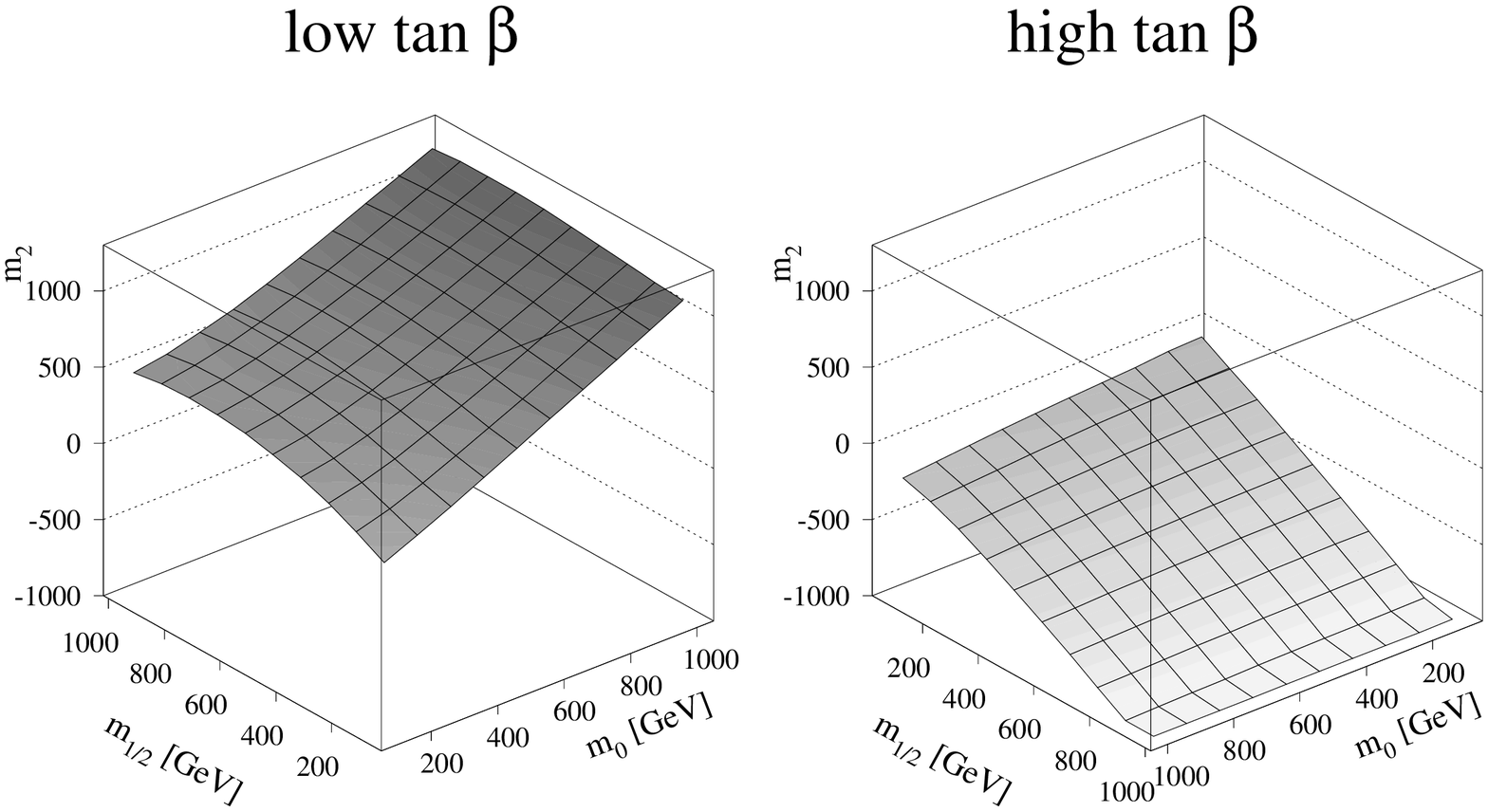}
\end{center}
\caption{\label{\figXIXb}The mass   $m_2$ in the Higgs potential 
  at $\mz$ (Born level) in GeV as function of $\mze$ and $\mha$ for
  the low and high $\tb$ scenario, respectively.  }
\end{figure*}

\begin{figure*}
\begin{center}
  \leavevmode
  \epsfxsize=15cm
  \epsffile{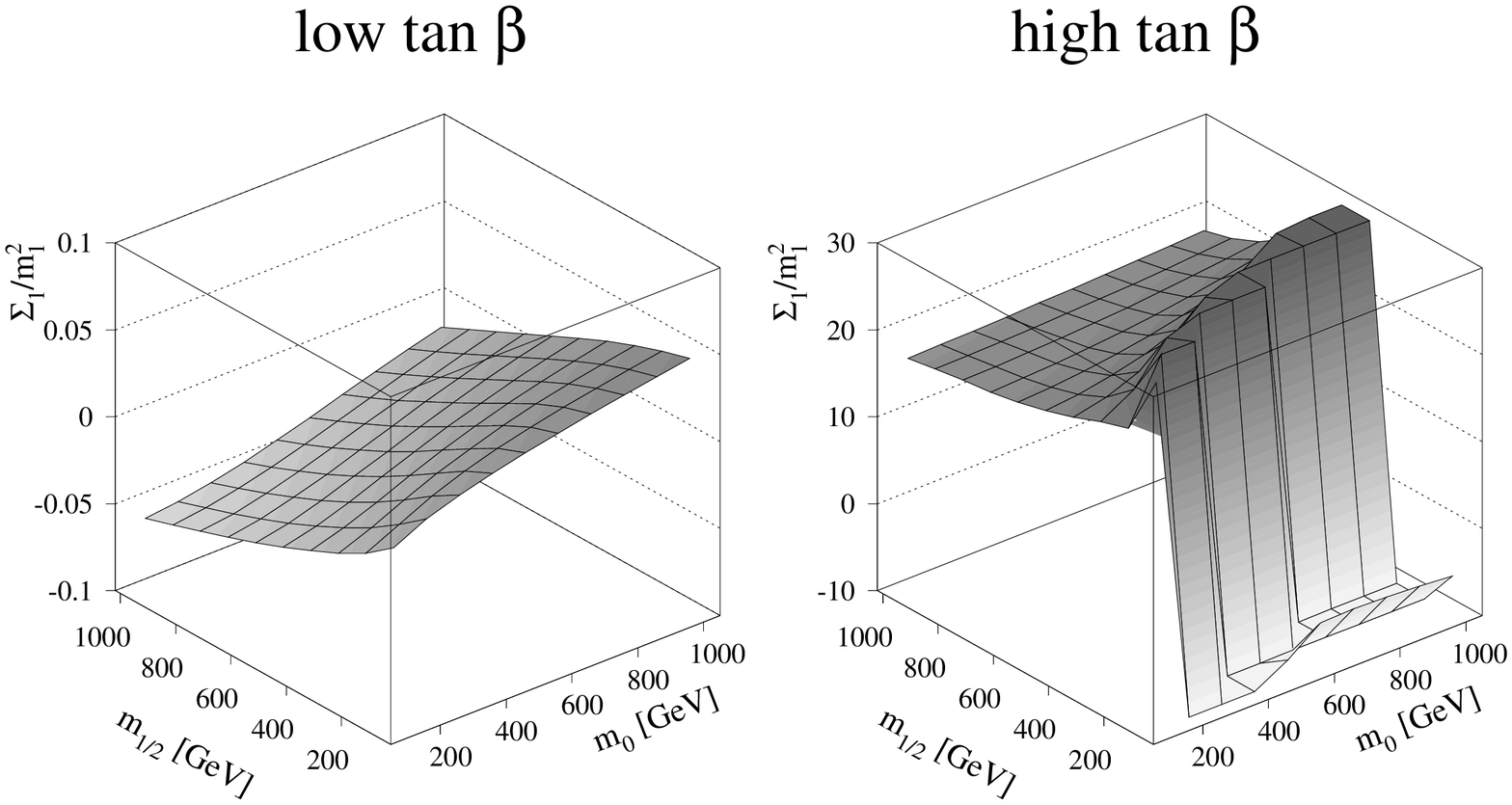}
\end{center}
\caption{\label{\figXIXc}The one-loop corrections  $\Sigma_1/m_1^2$ 
  at $\mz$ as function of $\mze$ and $\mha$ for the low and high $\tb$
  scenario, respectively.}
\end{figure*}

\begin{figure*}
\begin{center}
  \leavevmode
  \epsfxsize=15cm
  \epsffile{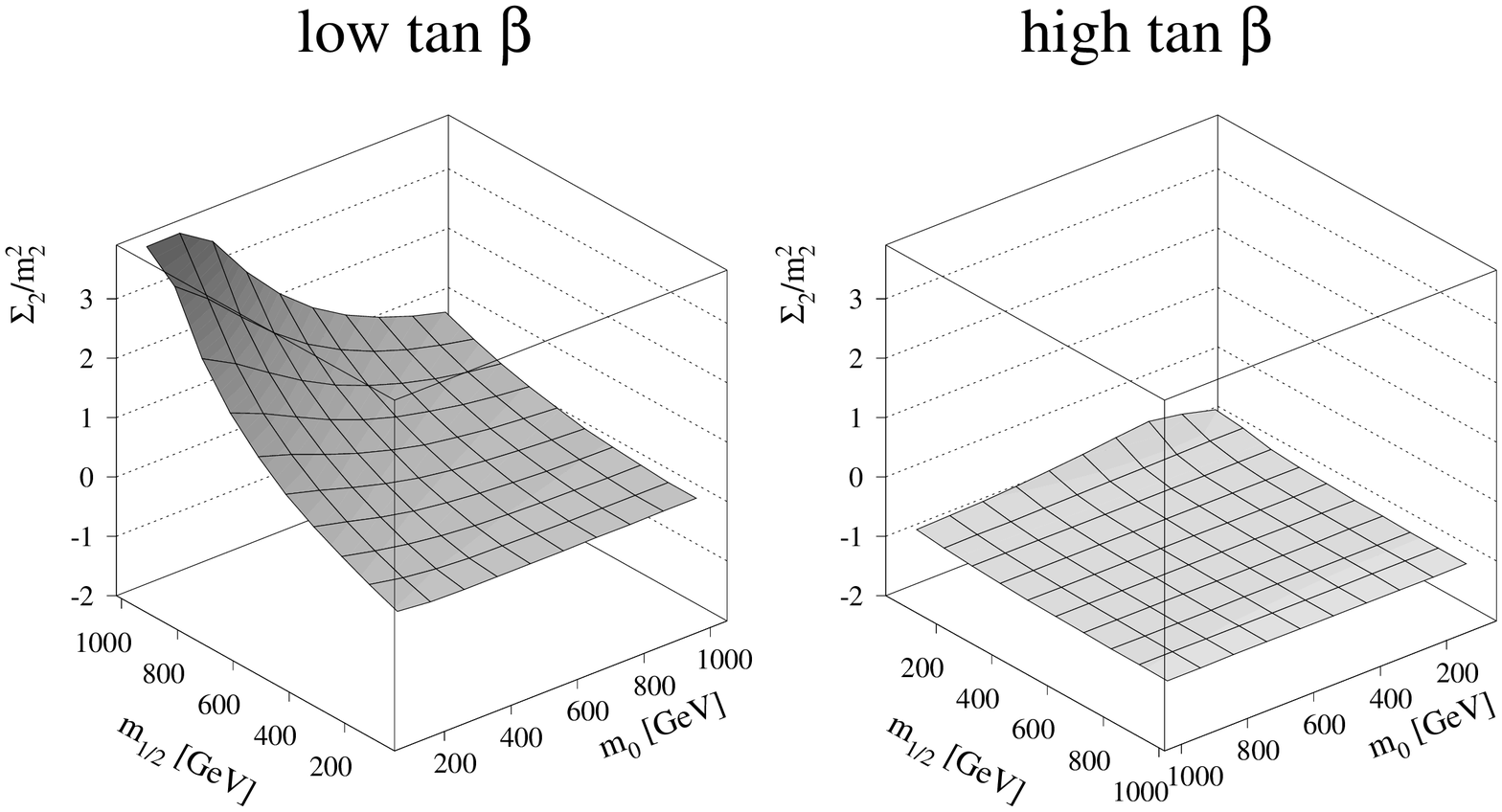}
\end{center}
\caption{\label{\figXIXd}The one-loop corrections  $\Sigma_2/m_2^2$ 
  at $\mz$ as function of $\mze$ and $\mha$ for the low and high $\tb$
  scenario, respectively.}
\end{figure*}

\begin{figure*}
\begin{center}
  \leavevmode
  \epsfxsize=15cm
  \epsffile{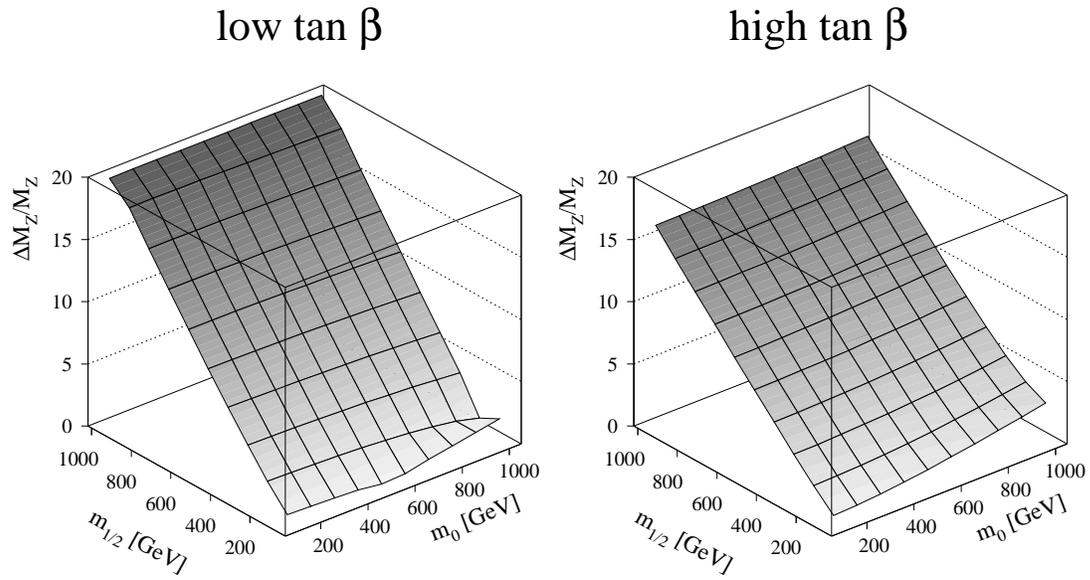}
\end{center}
\caption{\label{\figXIXe}The one-loop radiative corrections
  $\Delta(\mz)/\mz$ as function of $\mze$ and $\mha$ for the low and
  high $\tb$ scenario, respectively.}
\end{figure*}

For large values of $m_0$ and $m_{1/2}$ the masses of the
superpartners and the normal particles become different. In this case
the famous cancellation of quadratic divergencies in supersymmetry
does not work anymore, leading to large corrections for the Higgs
potential parameters  and electroweak scale, as shown in figs.
\ref{\figXIXa}-\ref{\figXIXe}.  For clarity the Born terms have been
displayed, too. 

It is a question of taste if one should exclude the regions with large
corrections. In our opinion such a fine tuning argument is difficult to
use for a mass scale below 1 TeV, and the whole region up to 1 TeV
should be considered, leading to quite large upper limits~in case of
the low $\tb$ scenario\cite{bek}.

\subsection{Discussion of the remaining constraints}

\begin{figure*}
 \begin{center}
  \leavevmode
  \epsfxsize=15cm
  \epsffile{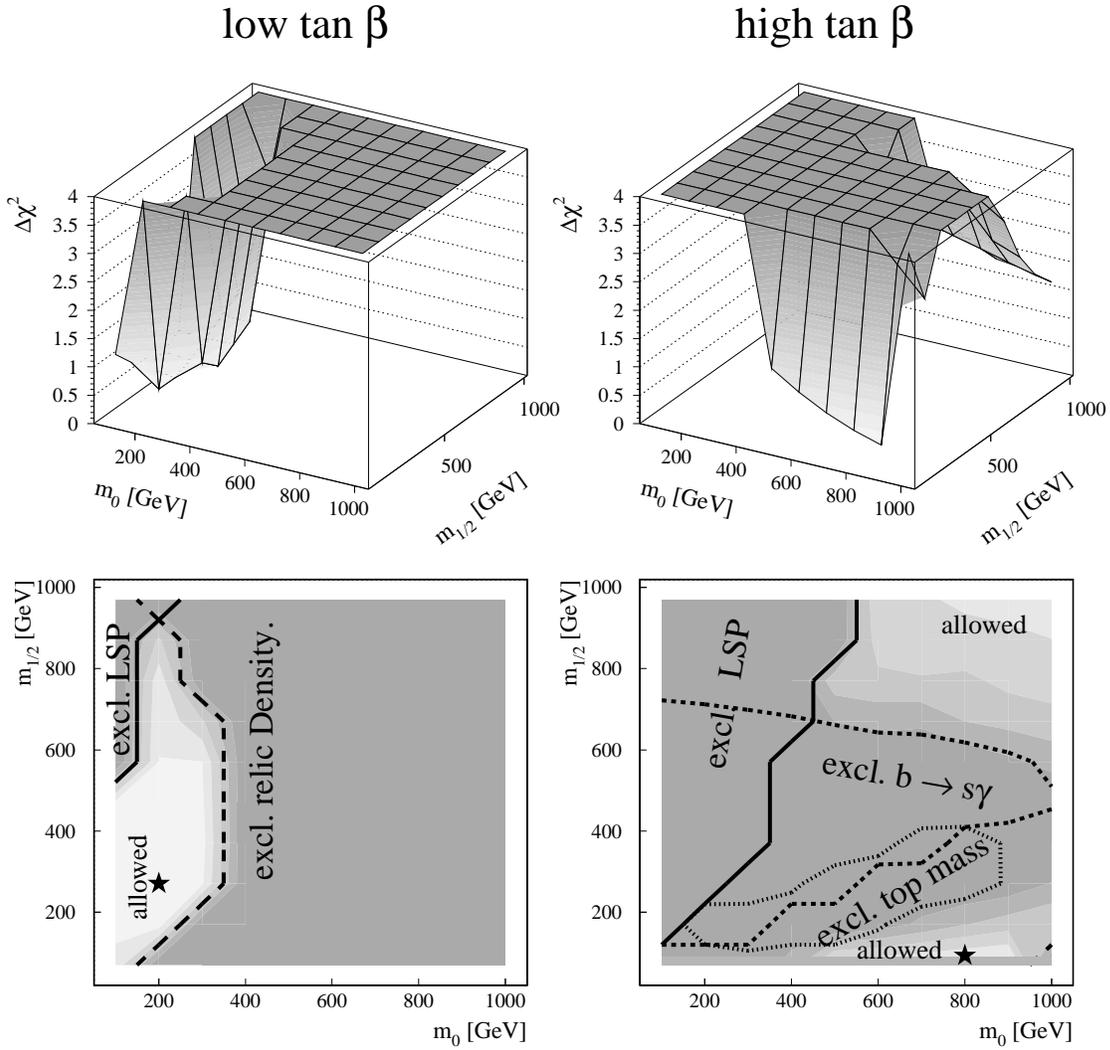}
\end{center}
\caption{\label{\figIII}The total $\chi^2$-distribution for 
  low and high $\tb$ solutions (top) as well as the projections
  (bottom). The different shades indicate steps of $\Delta\chi^2 = 1$,
  so basically only the light region is allowed. The stars indicate
  the optimum solution.}
\end{figure*}

In fig.~\ref{\figIII} the total $\chi^2$ distribution is shown as a
function of $\mze$ and $\mha$ for the two values of $\tb$ determined
above. One observes clear minima at $\mze,\mha$ around (200,270) and
(800,88), as indicated by the stars in the projections.  The different
shades correspond to $\Delta \chi^2$ steps of 1.  Note the sharp
increase in $\chi^2$, so basically only the light shaded regions are
allowed independent of the exact $\chi^2$ cut. The fitted values of
the other parameters are shown in table~\ref{t2a} and the
corresponding SUSY masses are given in table~\ref{t2}. 

\begin{figure*}
 \begin{center}
  \leavevmode
  \epsfxsize=15cm
  \epsffile{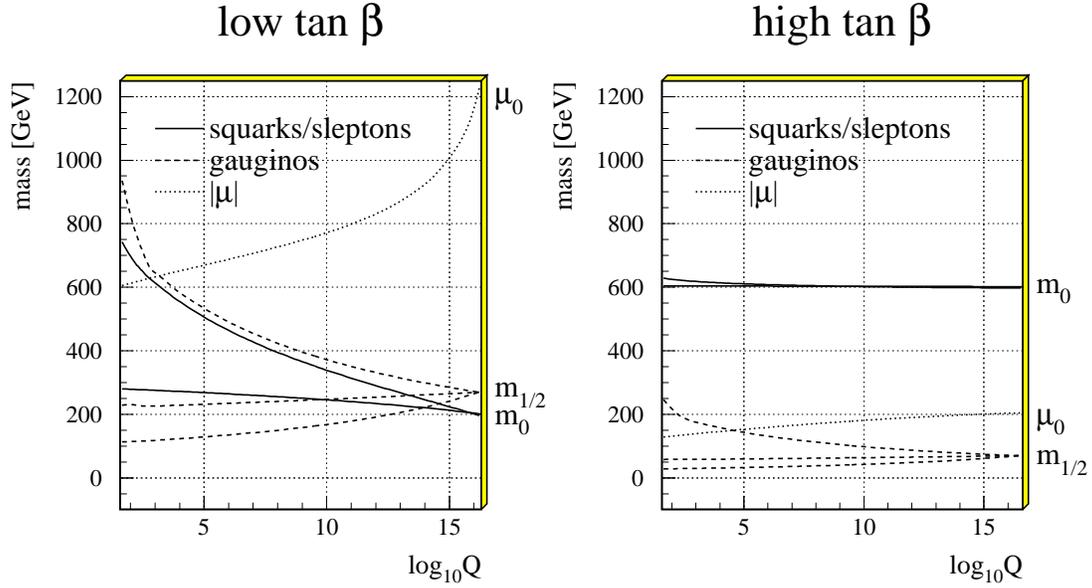}
\end{center}
\caption{\label{\figIV}The running of the particle masses and the
  $\mu$ parameter for low and high $\tb$ values.}
\end{figure*}

The running of some masses down to $M_Z$ is shown in
fig.~\ref{\figIV}. The values in table~\ref{t2} are not the values at
 $M_Z$, but at the physical mass $m_i$ for each particle, since the
running was stopped at $m_i$. 
 
\begin{figure*}
\begin{center}
  \leavevmode
  \epsfxsize=15cm
  \epsffile{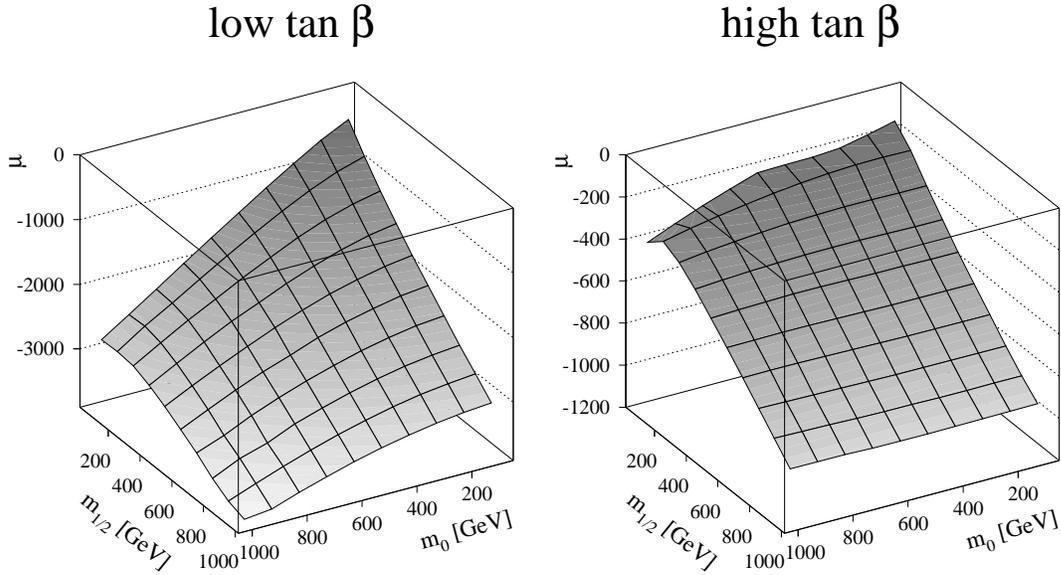}
\end{center}
\caption{\label{\figXVI}The Higgs mixing parameter $\mu_0$ at the GUT
  scale as function of $\mze$ and $\mha$ for the low and high $\tb$
  scenario, respectively.}
\end{figure*}

\begin{figure*}
 \begin{center}
  \leavevmode
  \epsfxsize=13cm
  \epsffile{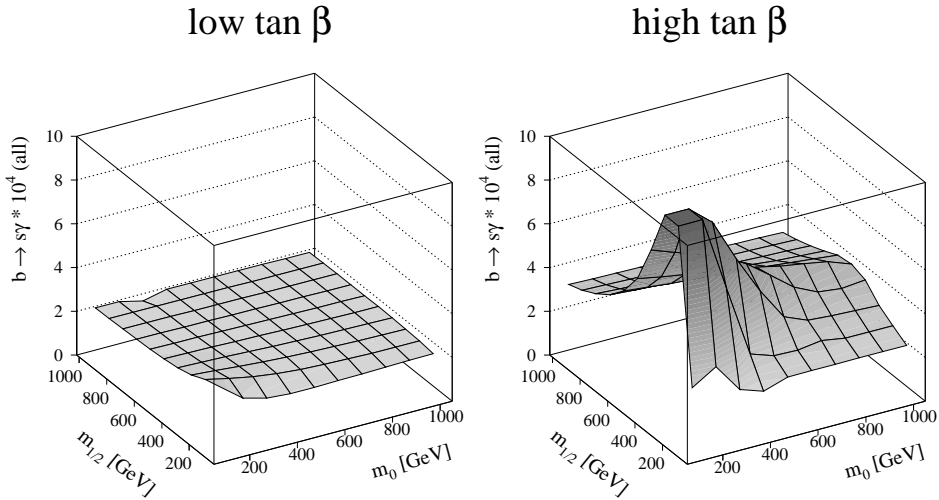}
\end{center}
\caption{\label{\figV}The branching ratio \bsg as function of $\mze$ and
  $\mha$. Note that for large \tb~only the region for $\mha < 120$ GeV
  yields values compatible with experimental results.}
\end{figure*}

\begin{figure*}
\begin{center}
  \leavevmode
  \epsfxsize=13cm
  \epsffile{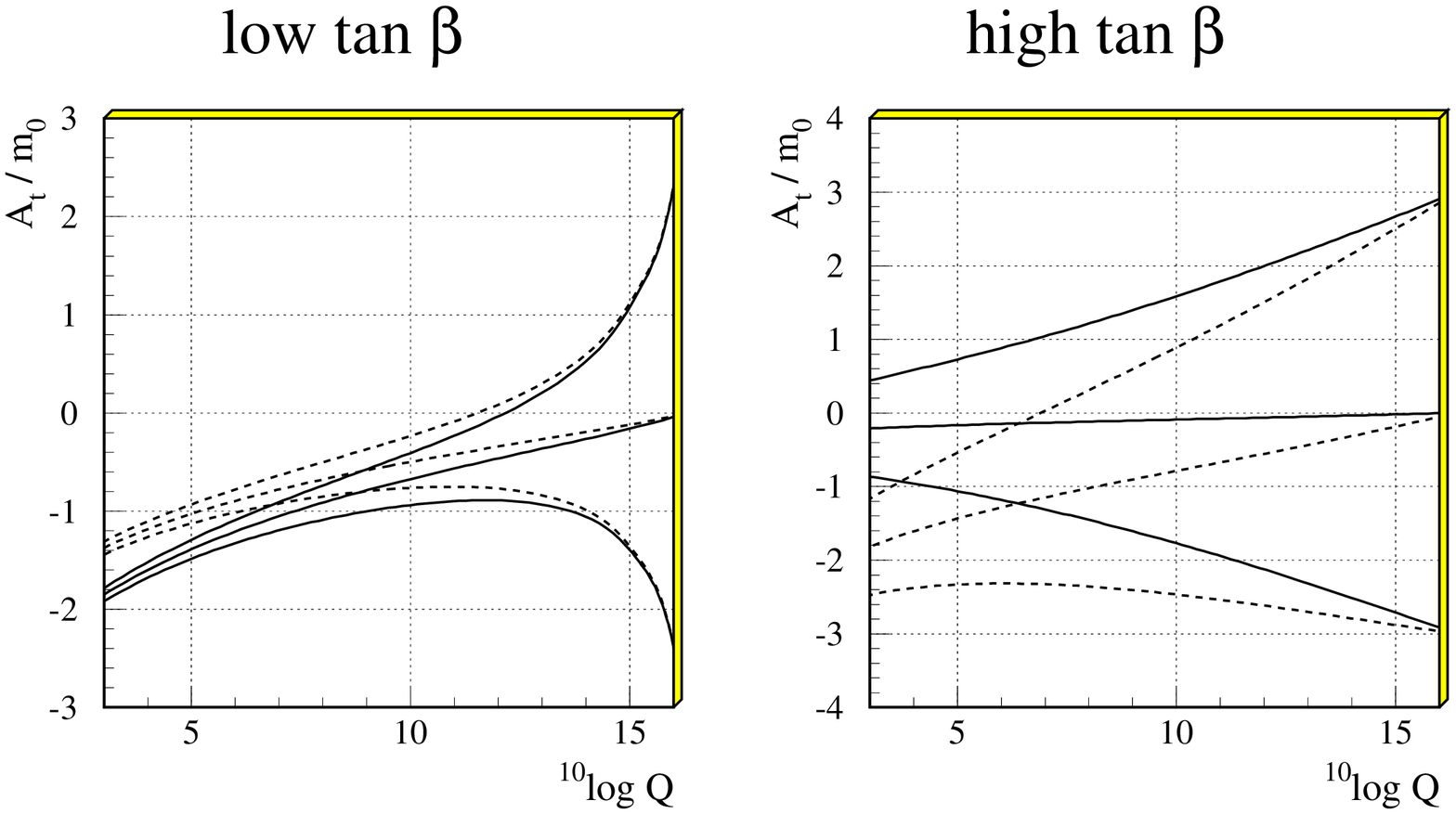}
\end{center}
\caption[]{\label{\figXa}The running of $A_t$.
  The values of ($\mze,\mha$) were chosen to be (200,270) and
  (800,88) GeV for the low and high $\tb$ scenario, respectively.
  However, the {\it fixed point} behaviour is found for other values
  of $\mze,\mha$ too: at low values of $\tb$ a strong convergence to a
  single value is found, while for high values this tendency is much
  less pronounced.}
\end{figure*}

\begin{figure*}
 \begin{center}
  \leavevmode
  \epsfxsize=13cm
  \epsffile{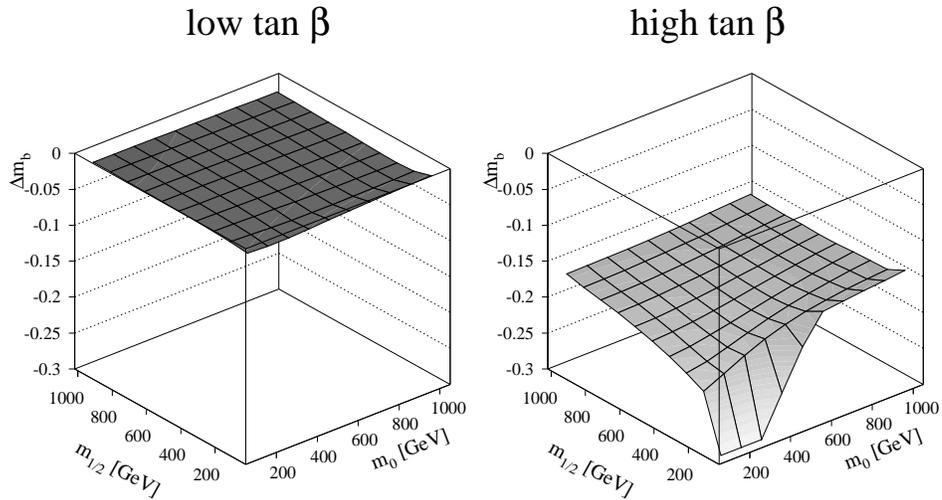}
\end{center}
\caption{\label{\figIX}Corrections to the bottom quark mass from
  gluino, charged Higgses and Higgsino loop contributions in the MSSM
  as function of $\mze$ and $\mha$. Note the large negative
  corrections for $|\mu|<0$ in this case. Positive $\mu$-values would
  yield a large positive contribution, which excludes bottom-$\tau$
  unification for most of the parameter space.}
\end{figure*}

\begin{figure*}
 \begin{center}
  \leavevmode
  \epsfxsize=15cm
  \epsffile{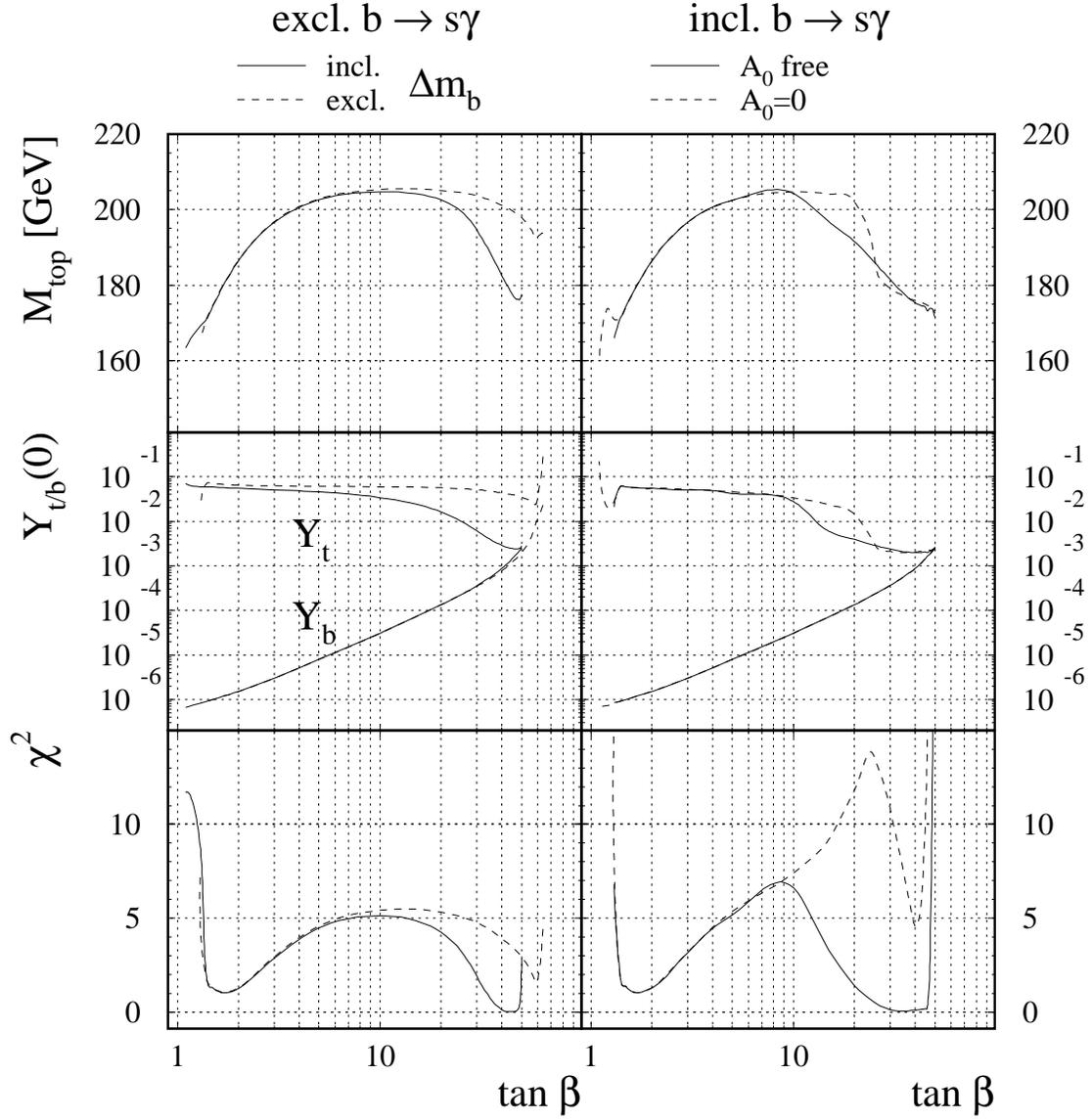}
\end{center}
\caption{\label{\figX}The top mass as function of $\tb$ for $\mze=600$
  and $\mha=70$ GeV. The various curves show the influence of the
  $\Delta m_b$ corrections, the \bsg branching ratio and the trilinear
  couplings of the third generation ($A_t=A_b=A_\tau=A_0$) at the GUT
  scale. These effects are only important for the higher values of
  $\tb$.}
\end{figure*}

\begin{figure*}
 \begin{center}
  \leavevmode
  \epsfxsize=15cm
  \epsffile{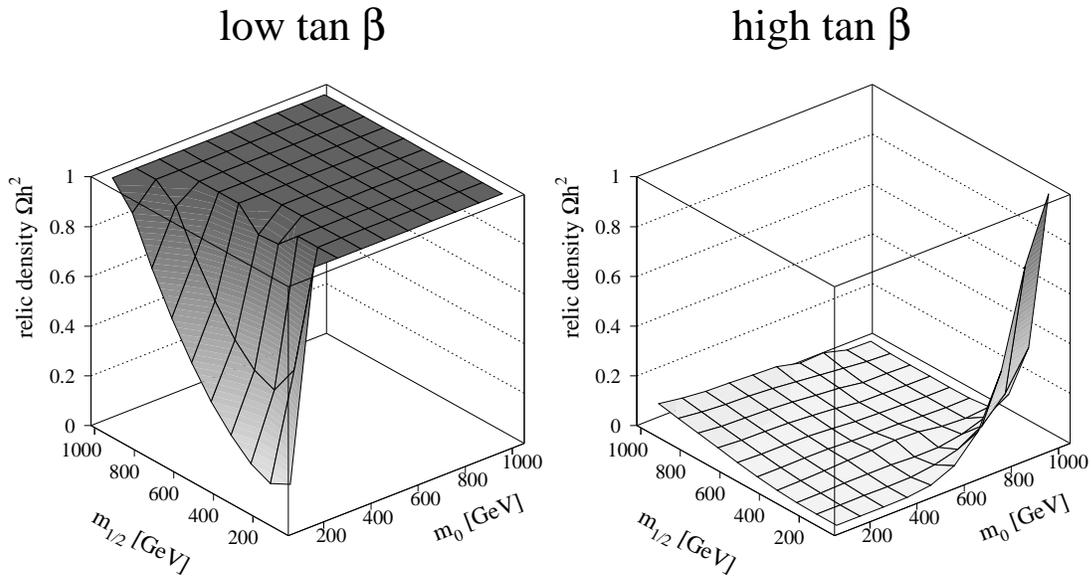}
\end{center}
\caption{\label{\figVI}The relic density as function of 
  $\mze$ and $\mha$ for the low and high $\tb$ scenario,
  respectively.}
\end{figure*}

The contours in fig.~\ref{\figIII} show the regions excluded by
different constraints used in the analysis, as will be discussed
below.
\begin{itemize}
\item \underline{\bf LSP Constraint:} The requirement that the LSP is
  neutral excludes the regions with small $m_0$ and relatively large
  $m_{1/2}$, since in this case one of the scalar staus becomes the LSP
  after mixing via the off-diagonal elements in the mass matrix
  (eq.~\ref{staumat}). The LSP constraint is especially effective at
  the high \tb region, since the off-diagonal element is proportional
  to $A_t m_0 - \mu\tan\beta$.
\item \underline{\bf \bsg Rate:} The predicted \bsg rate is shown in
  fig.~\ref{\figV} as function of $m_0$ and $m_{1/2}$ (with all other
  parameters optimized by the fit). At low \tb the \bsg rate is close
  to its SM value for most of the plane. The charginos and/or the
  charged Higgses are only light enough at small values of $m_0$ and
  $m_{1/2}$ to contribute significantly. The trilinear couplings
  were found to play a negligible role for low \tb. Varying
  them between $\pm 3\mze$ did not
  change the results significantly, since $A_t$ shows a {\it fixed
  point} behaviour in this case: its value at $\mz$ is practically
  independent of the starting value at the GUT scale, as shown in
  fig.~\ref{\figXa}.

  However, for large $\tb$ the trilinear coupling needs to be
  left free, since it is difficult to fit simultaneously $\bsg$, $m_b$
  and $m_\tau$. The reason is that the corrections to $m_b$ are
  large for large values of \tb due to the large contributions
  from $\tilde{g}-\tilde{q}$ and $\tilde{\chi}^\pm - \tilde{t}$
  loops proportional to $\mu\tb$ (see eq.~\ref{deltab}). They 
  become of the order of 10-20\%, as shown in fig.~\ref{\figIX}. In
  order to obtain $m_b(M_Z)$ as low as 2.84 GeV, these corrections
  have to be negative, thus requiring $\mu$ to be negative.
  
  As shown in fig.~\ref{\figV} the \bsg rate is too large in
  most of the parameter region for large \tb. In order to reduce
  this rate one needs $A_t(M_Z)>0$ for $\mu<0$ (see eq.~\ref{chargino}).
  Since for large \tb $A_t$ does not show a fix point behaviour
  (see fig.~\ref{\figXa}), this is possible.  The
  $\chi^2$ for large $\tb$ and $A_0 = 0$ is much worse than for fits
  in which $A_0$ is left free, as can be seen in fig.~\ref{\figX}; also 
  the influence of
  $\Delta m_b$ on the $b-\tau$ unification solution is shown on the
  left side.  Note that the $\Delta m_b$ corrections improve the fit.
  

\item \underline{\bf Relic Density:} the predicted Relic Density is
  shown in fig.~\ref{\figVI}. For the low \tb scenario the value of
  $\mu$ from EWSB is large (see fig.~\ref{\figXVI}). In this case
  there is little mixing between the Higgsino and Gaugino sector as is
  apparent from the neutralino mass matrix: for $|\mu| \gg M_1 \approx
  0.4 m_{1/2}$ the mass of the LSP is simply $0.4 m_{1/2}$ and the
  ``bino'' purity is 99\% (see table~\ref{t2a}). For the high \tb
  scenario $\mu$ is much smaller (see fig.~\ref{\figXVI}) and the
  Higgsino admixture becomes larger.  This leads to an enhancement of
  $\tilde\chi-\tilde\chi$ annihilation via the s-channel Z boson
  exchange, thus reducing the relic density.  As a result, in the
  large $\tb$ case the constraint $\Omega h_0^2 < 1$ is almost always
  satisfied unlike in the case of low $\tb$.
\end{itemize}

\begin{figure*}
\begin{center}
  \leavevmode
  \epsfxsize=15cm
  \epsffile{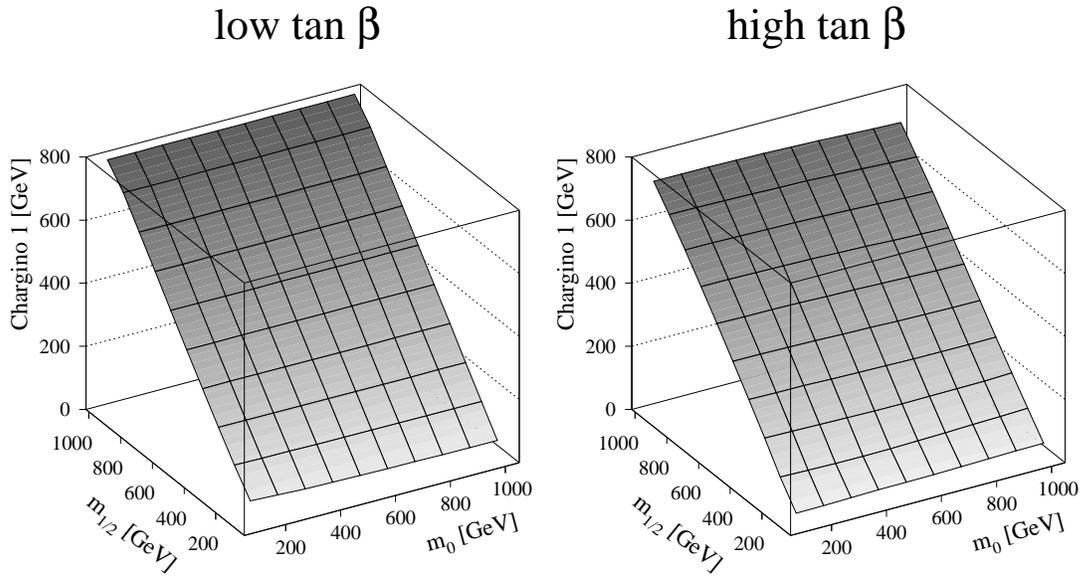}
\end{center}
\caption{\label{\figVII}The lightest chargino  mass as function of  
  $\mze$ and $\mha$ for the low and high $\tb$ scenario,
  respectively.}
\end{figure*}

The mass of the lightest chargino is about $0.7~-0.8~\mha$, as shown
in fig.~\ref{\figVII}.  
The low value of $\mha$ for the best fit at large $\tb$ is mainly
restricted by the chargino mass limit of $65GeV$ included in the fit.
However, it will be difficult to exclude the large $\tb$ scenario,
since a change of the limit on the chargino masses from 65~GeV to the
possible limit of 95~GeV at LEPII does not significantly change the
$\chi^2$ of the best fit.

Of course, this conclusion depends sensitively on the \Bbsg value. For
large $\mha$ values, the prediction for this branching ratio is only 2
or 3 standard deviations above its experimental value (see
fig.~\ref{\figV}).

\begin{figure*}
 \begin{center}
  \leavevmode
  \epsfxsize=15cm
  \epsffile{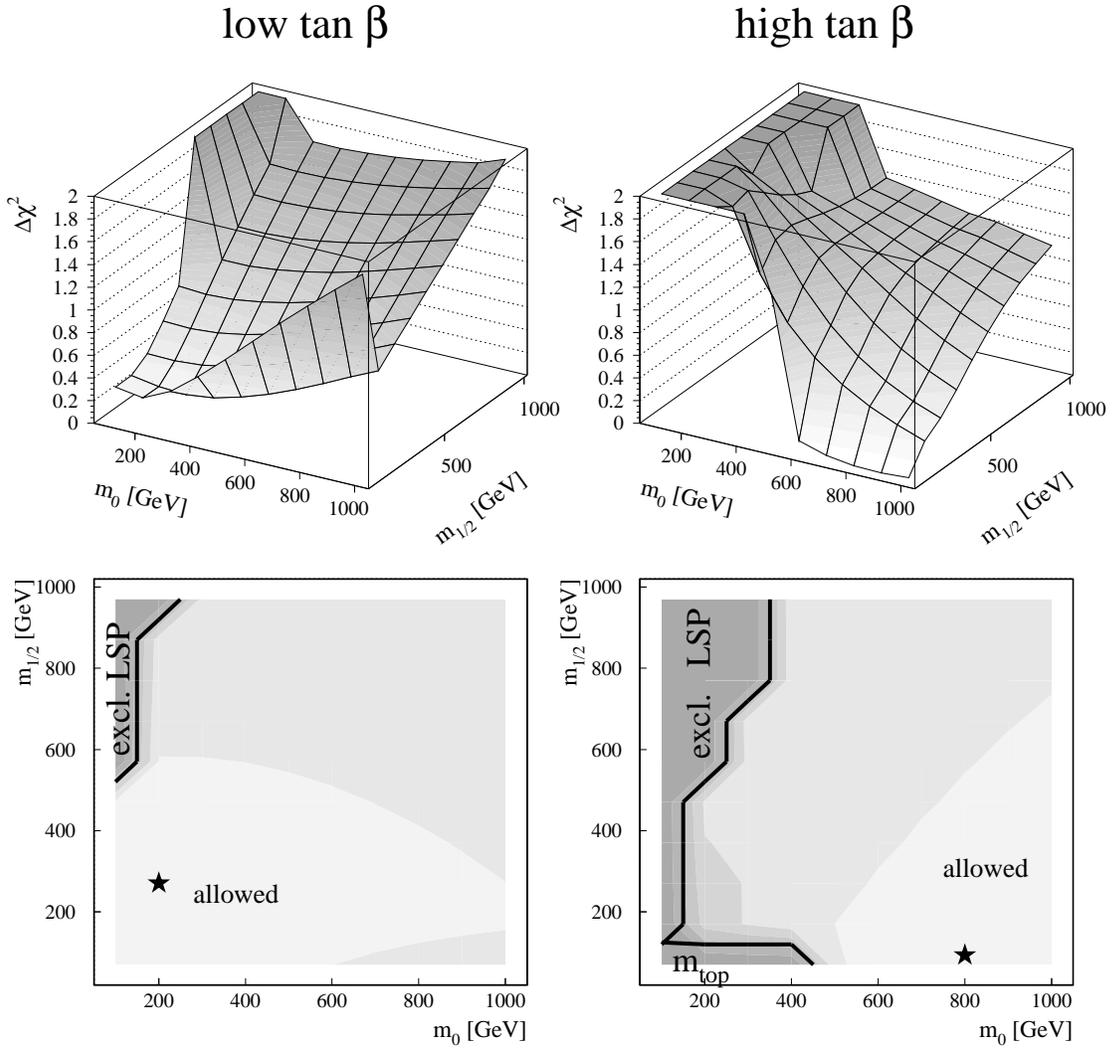}
\end{center}
\caption[]{\label{\figVIII}As fig.~\ref{\figIII}, but only including the
  constraints from unification and electroweak symmetry breaking.}
\end{figure*}

Without the constraints from \bsg and dark matter, large values of the
SUSY scale cannot be excluded, since the $\chi^2$ from gauge and
Yukawa coupling unification and electroweak symmetry breaking alone
does not exclude these regions (see fig.~\ref{\figVIII}). However,
there is a clear preference for the lighter SUSY scales.

\begin{figure*}
\begin{center}
  \leavevmode
  \epsfxsize=15cm
  \epsffile{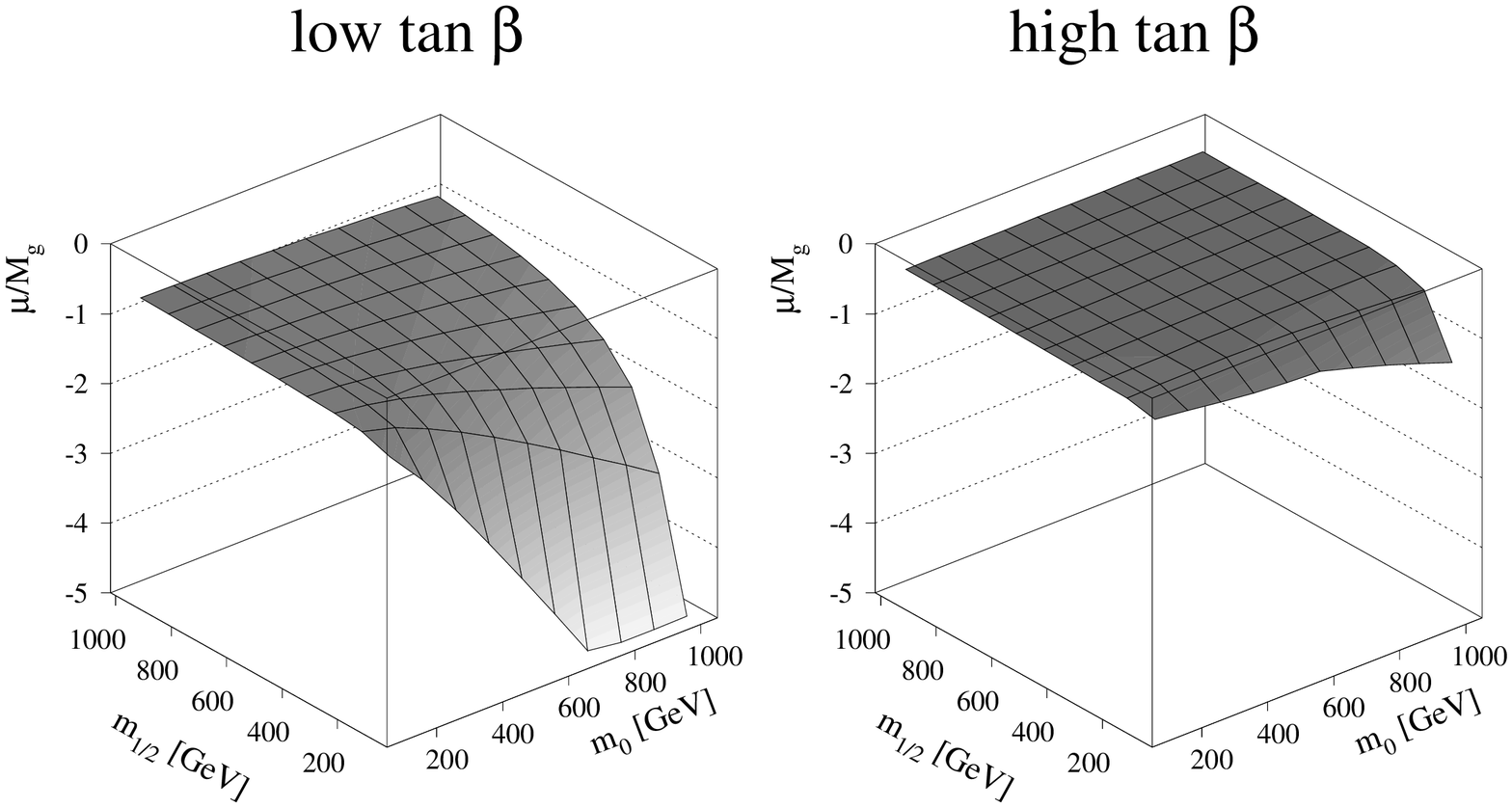}
\end{center}
\caption{\label{\figXII}The ratio of $\mu(\mz)$ and the gluino mass 
  as function of $\mze$ and $\mha$ for the low and high $\tb$
  scenario, respectively.}
\end{figure*}

\begin{figure*}
\begin{center}
  \leavevmode
  \epsfxsize=15cm
  \epsffile{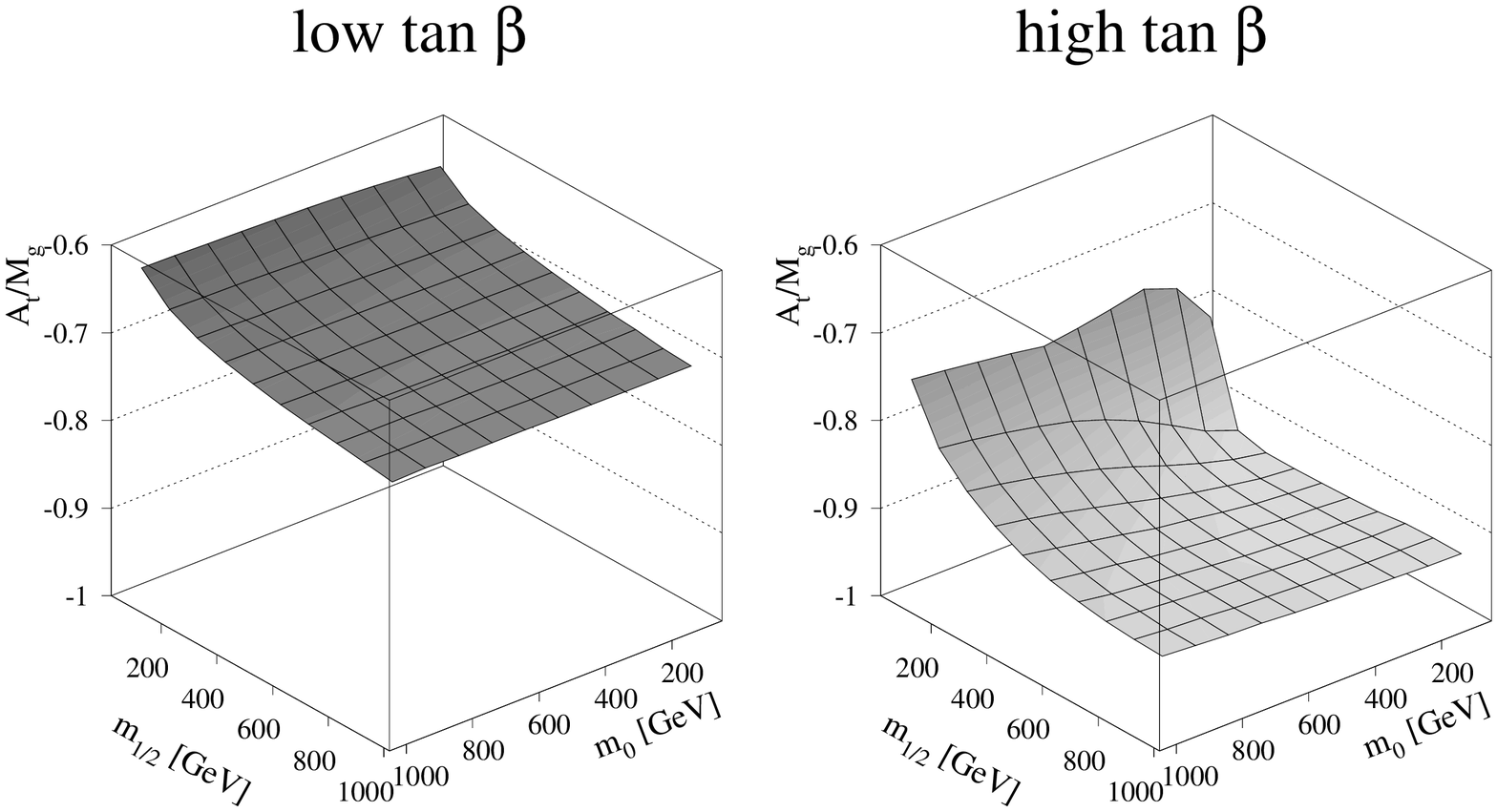}
\end{center}
\caption{\label{\figXIII}The ratio of $A_t$ and the gluino mass as 
  function of $\mze$ and $\mha$ for the low and high $\tb$ scenario,
  respectively ($A_0=0$).}
\end{figure*}

\begin{figure*}
\begin{center}
  \leavevmode
  \epsfxsize=15cm
  \epsffile{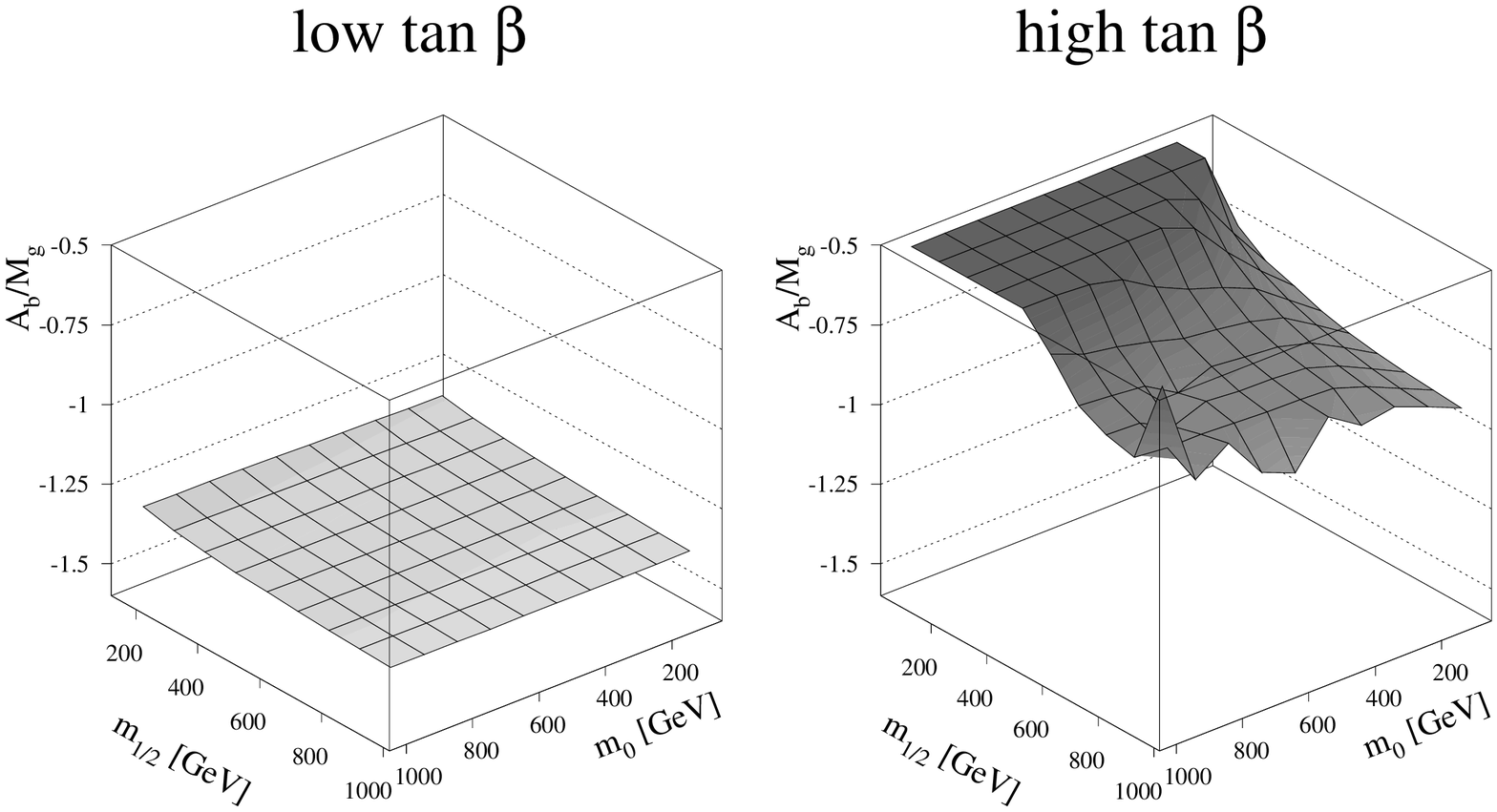}
\end{center}
\caption{\label{\figXIV}The ratio of $A_b(\mz)$ and the gluino mass as
  function of $\mze$ and $\mha$ for the low and high $\tb$ scenario,
  respectively ($A_0=0$).}
\end{figure*}

\begin{figure*}
\begin{center}
  \leavevmode
  \epsfxsize=15cm
  \epsffile{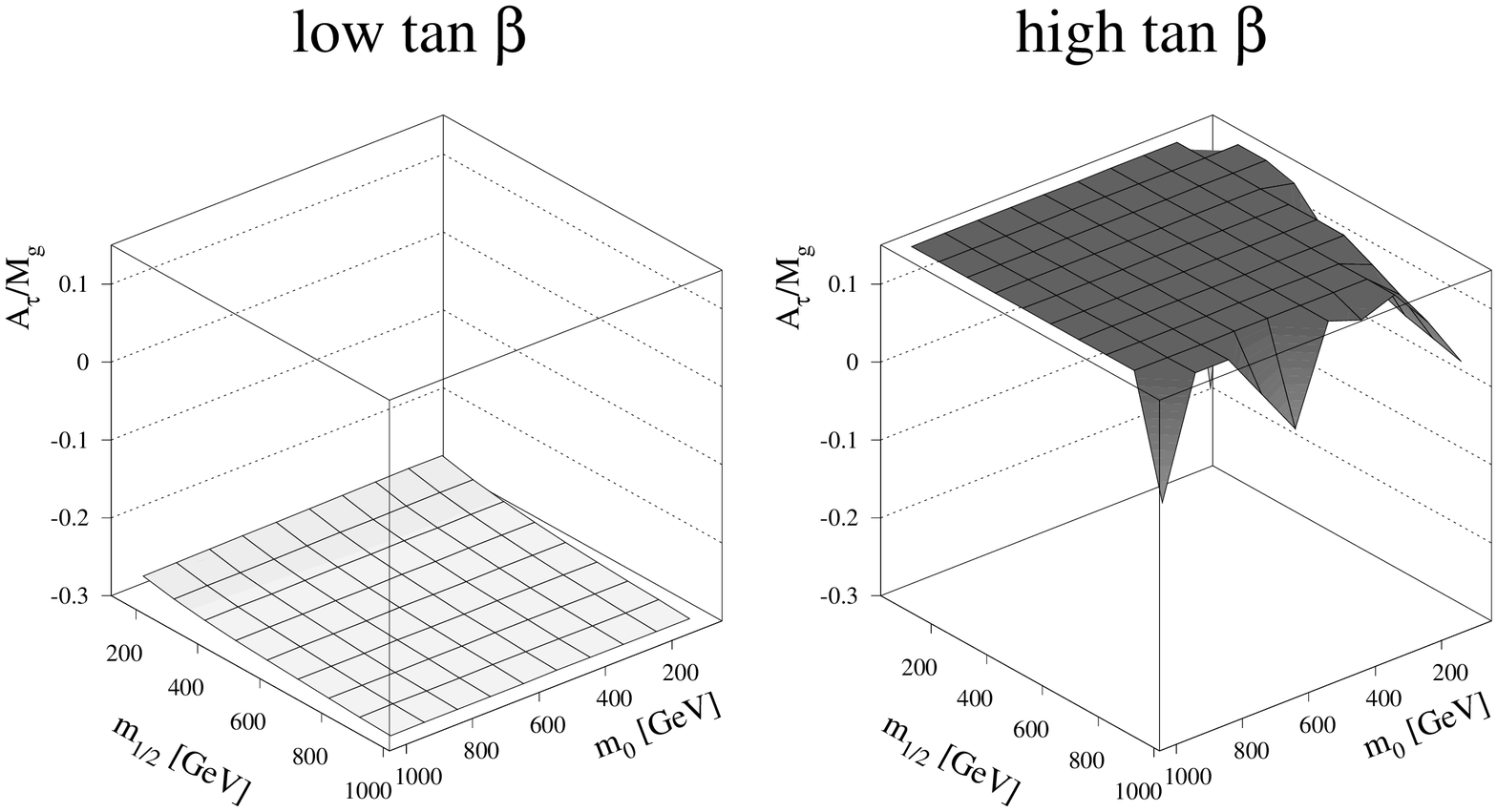}
\end{center}
\caption{\label{\figXV}The ratio of $A_\tau(\mz)$ and the gluino mass
  as function of $\mze$ and $\mha$ for the low and high $\tb$
  scenario, respectively ($A_0=0$).}
\end{figure*}

\begin{figure*}
\begin{center}
  \leavevmode
  \epsfxsize=15cm
  \epsffile{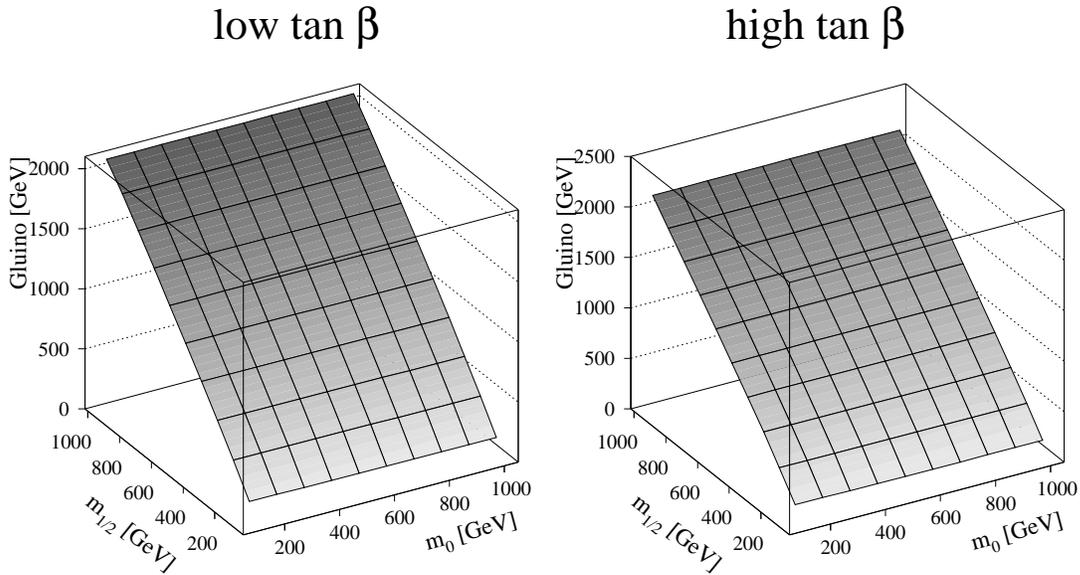}
\end{center}
\caption{\label{\figXI}The gluino mass as function of $\mze$ and $\mha$ 
  for the low and high $\tb$ scenario, respectively.}
\end{figure*}

The fitted values of the trilinear couplings and the Higgs mixing
parameter $\mu$ are strongly correlated with $\mha$, so the ratio of
these parameters at the electroweak scale and the gluino mass is
relatively constant and largely independent of $\mze$ (see figs.
\ref{\figXII} - \ref{\figXV}).  The gluino mass depends only on
$m_{1/2}$ ($M_{\tilde g}\approx 2.7~\mha$), as shown in
fig.~\ref{\figXI}.  Note from the figures that although the trilinear
couplings $A_t$, $A_b$ and $A_\tau$ have equal values at the GUT
scale, they are quite different at the electroweak scale due to the
different radiative corrections.

\section{Discovery Potential  at LEP II}
Table \ref{t2} shows that charginos, neutralinos and the lightest
Higgs belong to the lightest particles in the MSSM.

\begin{figure*}
\begin{center}
  \leavevmode
  \epsfxsize=15cm
  \epsffile{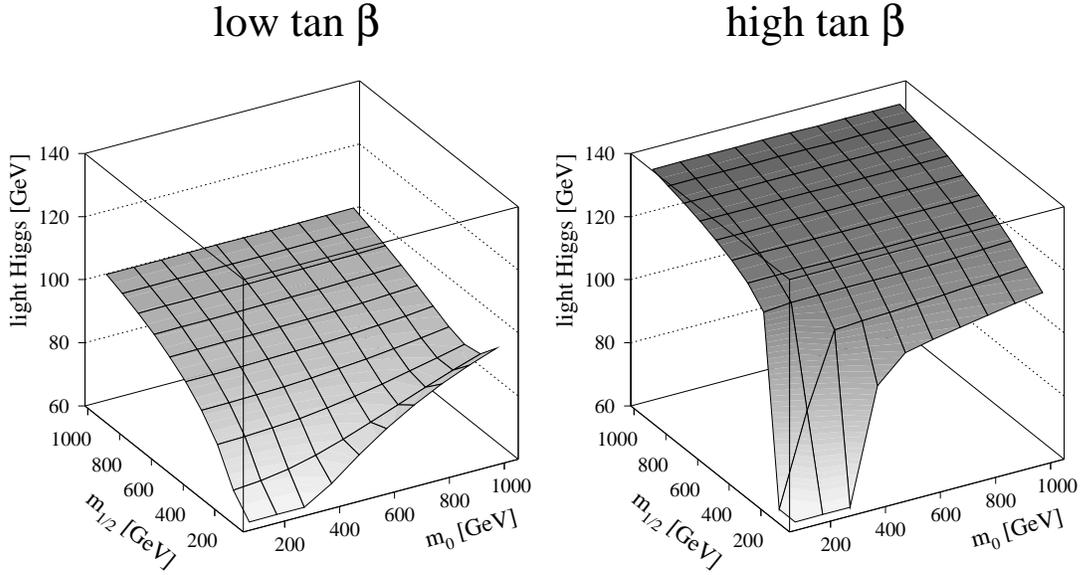}
\end{center}
\caption{\label{\figXVII}The mass of the lightest CP-even Higgs as
  function of $\mze$ and $\mha$ for the low and high $\tb$ scenario,
  respectively.  The sign of $\mu $ is negative, as required for the
  high $\tb$ solution, but chosen negative for low $\tb$.
  }
\end{figure*}

\begin{figure*}
\begin{center}
  \leavevmode
  \epsfxsize=15cm
  \epsffile{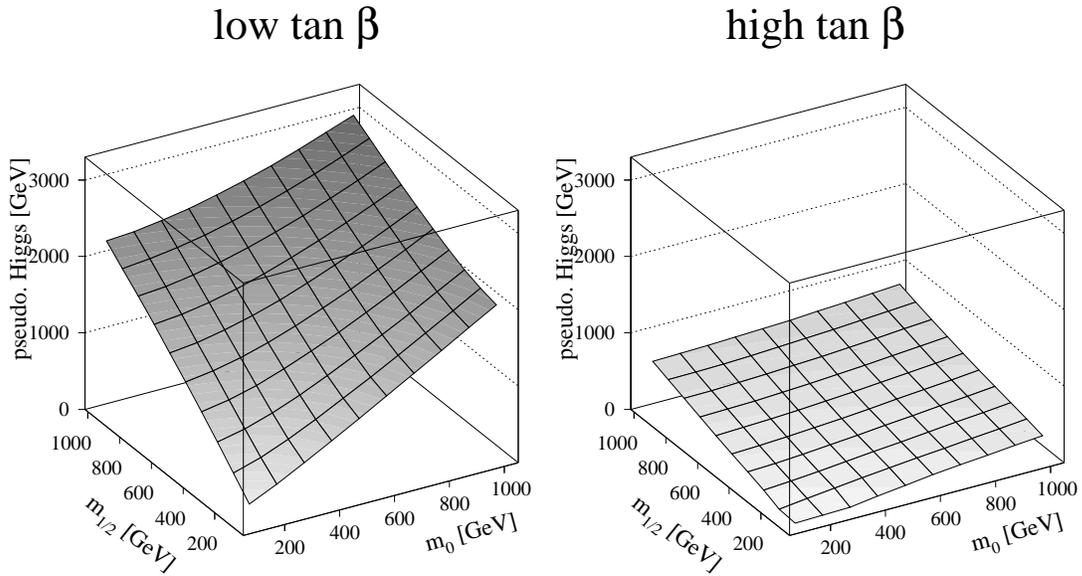}
\end{center}
\caption{\label{\figXVIII}The mass of the CP-odd Higgs as function of  
  $\mze$ and $\mha$ for the low and high $\tb$ scenario,
  respectively.}
\end{figure*}

In figs.  \ref{\figXVII},\ref{\figXVIII} the masses of the lightest
CP-even and CP-odd Higgs bosons are shown for the whole parameter
space for negative $\mu$-values.  At each point a fit was performed to
obtain the best solution for the GUT parameters.  The mass of the
lightest Higgs saturates at 100 GeV. For positive $\mu$-values and low
$\tb$ the maximum Higgs mass increases to 115 GeV.

For high $\tb$ only negative $\mu$-values are allowed, since positive
$\mu$-values yield a too high $b$-mass due to the large positive
corrections in that case, as discussed above.
   
The programs SUSYGEN~\cite{susygen} and special SUSY routines
\cite{isajetee} in ISAJET~\cite{isajet} have been used to calculate
the production cross sections.

\begin{figure*}
\begin{center}
  \leavevmode
  \epsfxsize=15cm
  \epsffile{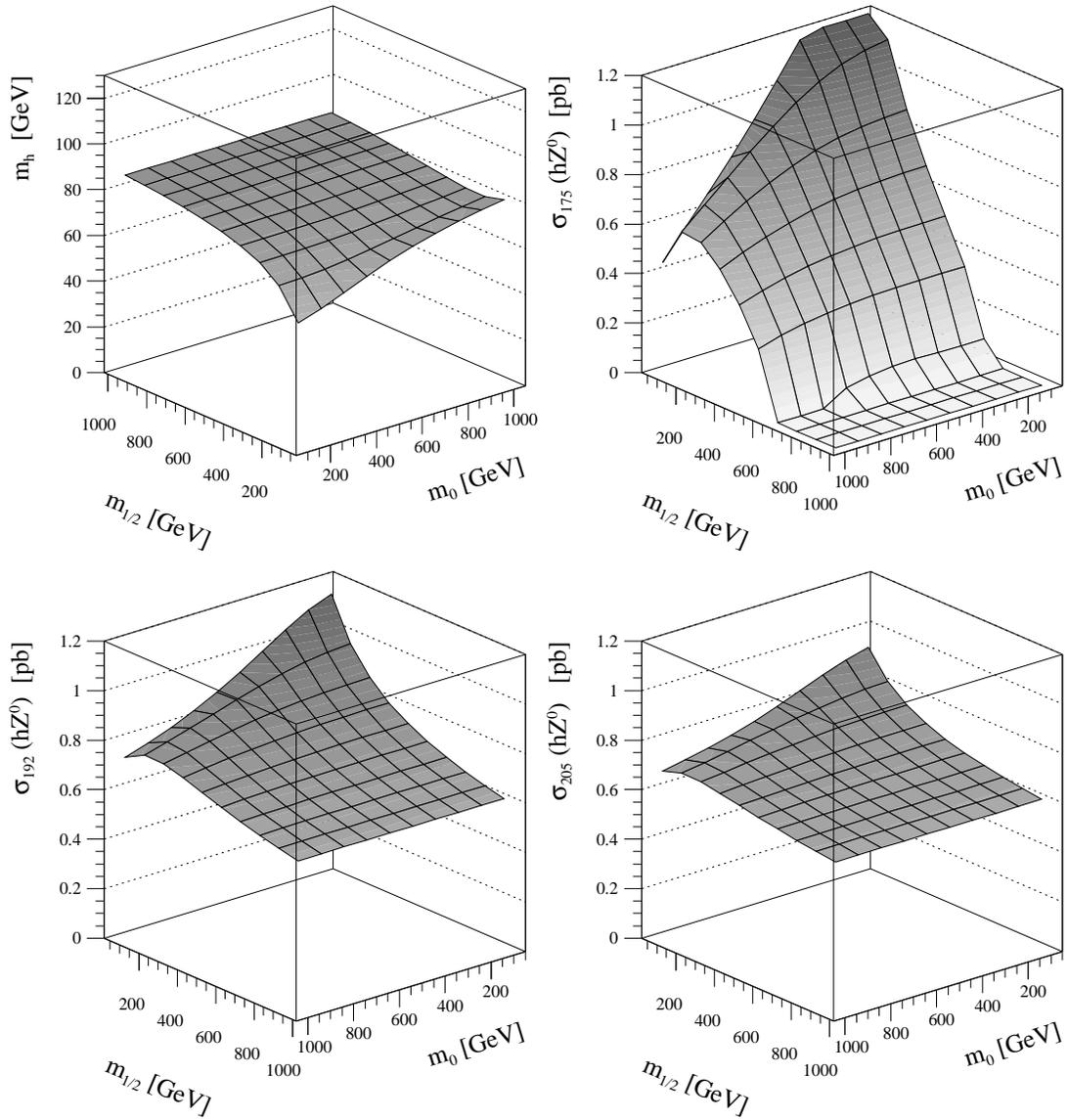}
\end{center}
\caption{\label{\figXX}The Higgs mass as function of $\mze$ and $\mha$
  for positive values of $\mu$ and $\tb = 1.7$ (left top corner) and
  the main production cross sections for three different LEP II
  energies (175, 192 and 205 GeV). For negative $\mu$-values the cross
  sections are about 50\% larger.}
\end{figure*}

Fig.~\ref{\figXX} shows the mass of the lightest Higgs boson and the
corresponding Higgs production cross sections at three LEP energies as
functions of $m_0$ and $m_{1/2}$ for $\tan\beta=1.7$ and $m_t=180$ GeV. 

\begin{figure*}
\begin{center}
  \leavevmode
  \epsfxsize=15cm
  \epsffile{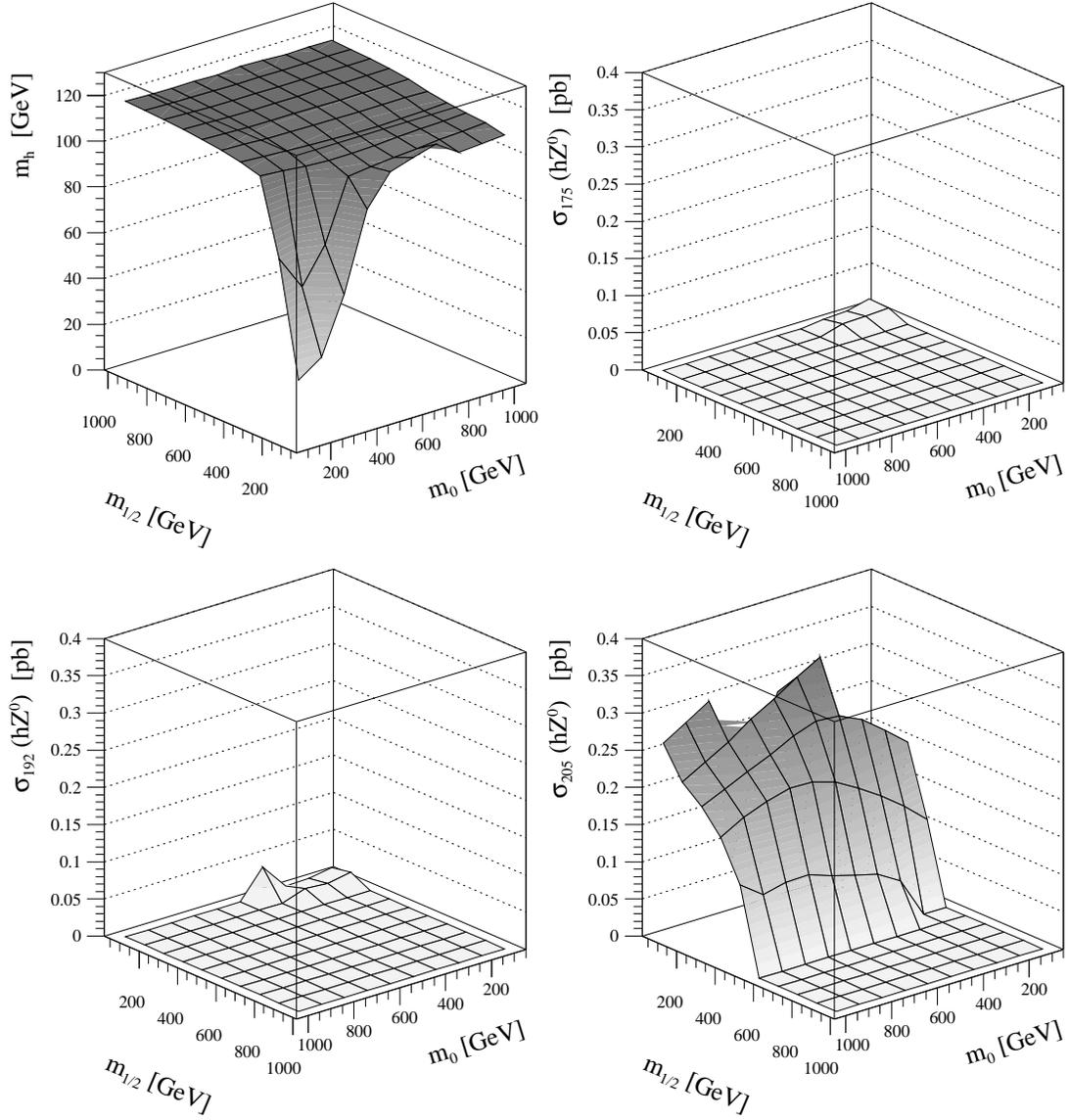}
\end{center}
\caption{\label{\figXXI}The Higgs mass as function of $\mze$ and
  $\mha$ for $\mu = -300$ GeV and $\tb = 46$ (left top corner) and the
  main production cross sections for three different LEP II energies
  (175, 192 and 205 GeV).}
\end{figure*}

Here the most significant second order corrections to the Higgs mass
have been incorporated \cite{ll}, which reduces the Higgs mass by
about 15 GeV \cite{bekhiggs}. In this case the foreseen LEP energy of
192 GeV is sufficient to cover the whole parameter space for the low
\tb scenario, provided the top mass is below 190 GeV. For the large
\tb scenario one needs the maximum possible LEP II energy of 205 GeV
in order to cover at least some part of the parameter space, as shown
in fig.~\ref{\figXXI}.

\begin{figure*}
\begin{center}
  \leavevmode
  \epsfxsize=15cm
  \epsffile{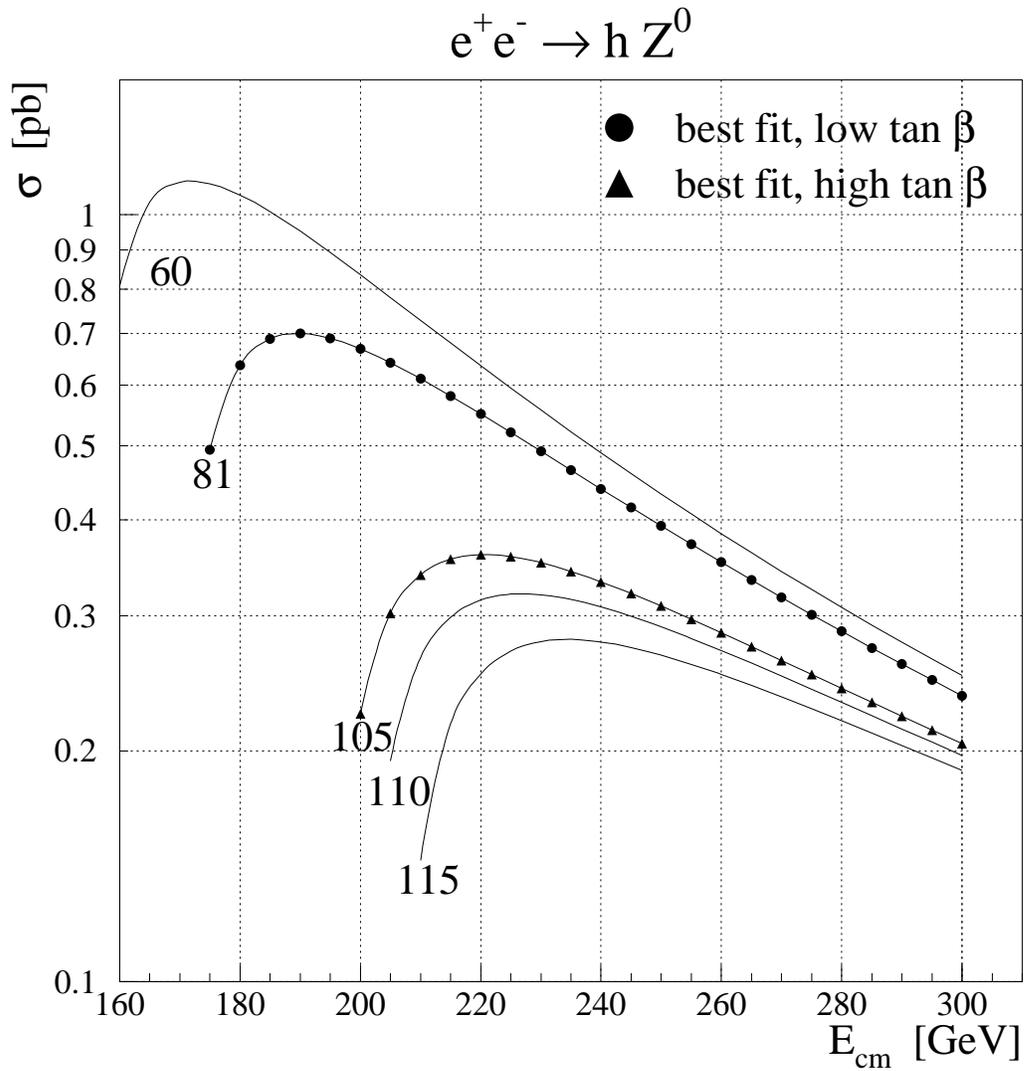}
\end{center}
\caption{\label{\figXXII}The cross section as function of the centre
  of mass energy for different Higgs masses, as indicated by the
  numbers (in GeV).}
\end{figure*}

\begin{figure*}
\begin{center}
  \leavevmode
  \epsfxsize=15cm
  \epsffile{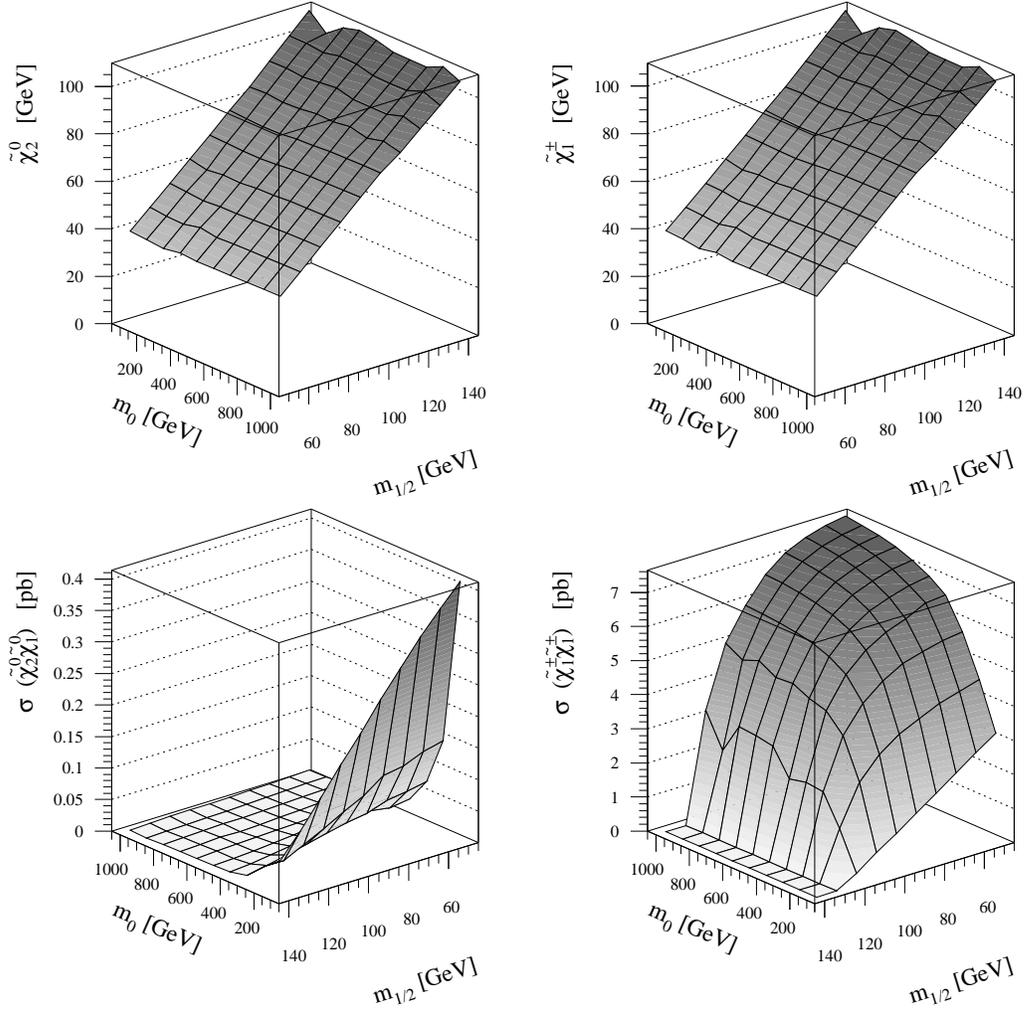}
\end{center}
\caption{\label{\figXXIII}The masses of the second lightest
  neutralino and chargino masses as well as the production cross
  sections for $\mu=-300$ GeV and $\tb = 46$. $\mu$ was kept to a
  representative value, since in most of the region the fit gave an
  unacceptable $\chi^2$, so $\mu$ could not be determined.  Positive
  values of $\mu$ give similar results.  The steep decrease in the
  chargino cross section at small values of $\mze$ is due to the light
  sneutrino, which leads to a strong negative interference between s-
  and t-channel in that case.  Fortunately, the neutralino production
  is large in that region,as shown by the plot in the left bottom
  corner.  }
\end{figure*}

The cross section dependence on the centre of mass energy is shown for
some representative Higgs masses in fig~\ref{\figXXII}. Clearly the
large \tb scenario is not very promising for the Higss mass discovery
at LEP II. However, in that case the discovery potential of the
chargino and neutralino searches is high, since the cross sections are
large for $m_{1/2}< 120$ GeV, as shown in fig.~\ref{\figXXIII}. This is
exactly the region allowed by all other constraints for large \tb (see
fig.~\ref{\figIII}), if one ignores the other solution with $m_{\tilde
  q}\ge 3$ TeV, i.e. $m_{1/2}\approx m_0 \approx 1$ TeV.

\section{Summary}
In the Constrained Minimal Supersymmetric Model (CMSSM) the optimum
values of the GUT scale parameters and the corresponding SUSY mass
spectra for the low and high $\tb$ scenario have been determined from
a combined fit to the low energy data on couplings, quark and lepton
masses of the third generation, the electroweak scale $\mz$, $\bsg$, and
the lifetime of the universe.
  
The solutions preferred by the best fit predict new particles at LEP
II: The lightest Higgs boson in case of the low \tb scenario and
chargino or neutralino production at the high \tb scenario. The upper
limits on the masses of these particles  are outside the LEP II
domain, but within reach of the LHC.

\section{Acknowledgement}
The research described in this publication was made possible in part
by support from the Human Capital and Mobility Fund (Contract
ERBCHRXCT 930345) from the European Community, and by support from the
German Bundesministerium f\"ur Bildung und Forschung (BMBF) (Contract
05-6KA16P) and from the Deutsche Forschungs-Gemeinschaft (DFG) for the
Graduiertenkolleg  in Karlsruhe.
\clearpage

\begin{table*}[p]
\renewcommand{\arraystretch}{1.30}
\renewcommand{\rb}[1]{\raisebox{1.75ex}[-1.75ex]{#1}}
\begin{center}
\normalsize
\begin{tabular}{|c|r|r|}
\hline
 \multicolumn{3}{|c|}{ Fitted SUSY parameters }                       \\
\hline
\hline
Symbol & \makebox[3.0cm]{\bf{low $\tb$}} & \makebox[3.0cm]{\bf{high $\tb$}}\\
\hline                                       
 $m_0$          &  200          &  800 \\    
\hline
 $m_{1/2}$      &  270          &  88  \\    
\hline
 $\mu(0)$       & -1084         &  -270\\    
\hline
 $\mu(\mz)$     &  -546         &  -220\\    
\hline
 $\tan\beta$    & 1.71          &  41.2\\    
\hline
 $Y_t(\mt)$     &  0.0080       &  0.0051\\  
\hline
 $Y_t(0)$       & 0.0416        &  0.0014\\  
\hline
 $Y_b(0)$       & 0.12E-05      &  0.0011\\  
\hline
$M_t^{pole}$    & 177           &  165 \\    
\hline
$m_t^{running}$ & 168           &  157\\     
\hline
 $1/\alpha_\rG$ & 24.8          &  24.3\\    
\hline
 $\MG$          & $1.6\;10^{16}$&  $2.5\;10^{16}$\\ 
\hline
\hline
 $A(0)m_0$     & 0             &  1256\\    
\hline
 $A_t(\mz)m_0$ &    -446       &  149 \\    
\hline
 $A_b(\mz)m_0$ &    -886       &  190 \\    
\hline
 $A_\tau(\mz)m_0$&  -182       &  639 \\    
\hline
 $m_1(\mz)m_0$     &    612        &  207 \\    
\hline
 $m_2(\mz)m_0$     &    262        &  -180 \\   
\hline
\end{tabular} \end{center}
\caption[]{\label{t2a}Values of the fitted SUSY parameters 
             for low and high $\tb$ (in GeV, when applicable).   
             The scale is  either $\mz$, $\mt$,  or $\rG$, as indicated
             in the first column by ($\mz$), ($\mt$) or (0), respectively. 
             The SUSY mass spectrum corresponding to these parameters
             is given in table \ref{t2}.}    
\end{table*}

\begin{table*}[p]
\renewcommand{\arraystretch}{1.30}
\renewcommand{\rb}[1]{\raisebox{1.75ex}[-1.75ex]{#1}}
\begin{center}
\normalsize
\begin{tabular}{|c|r|r|}
\hline
 \multicolumn{3}{|c|}{SUSY masses in [GeV]}             \\
\hline
\hline
Symbol & \makebox[3.0cm]{\bf{low $\tb$}} & \makebox[3.0cm]{\bf{high $\tb$}}\\
\hline
\hline
 $\tilde{\chi}^0_1(\tilde{B})$ &  116    & 35\\  
\hline                                        
 $\tilde{\chi}^0_2(\tilde{W}^3)$&  231   & 65\\  
\hline
 $\tilde{\chi}^{\pm}_1(\tilde{W}^\pm)$ &  231  & 65\\ 
\hline
  $\tilde{g}$                   &  658  & 236\\  
\hline  \hline                                
  $\tilde{e}_L$                 &  278  & 804\\  
\hline                                        
  $\tilde{e}_R$                 &  228  & 802\\  
\hline                                        
  $\tilde{\nu}_L$               &  273  & 799\\  
\hline  \hline                                
  $\tilde{q}_L$                 &  628  & 825\\  
\hline                                        
  $\tilde{q}_R$                 &  605  & 823\\  
\hline                                        
  $\tilde{\tau}_1$              &  227  & 557\\  
\hline                                        
  $\tilde{\tau}_2$              &  228  & 695\\  
\hline                                        
  $\tilde{b}_1$                 &  560  & 463\\  
\hline                                        
  $\tilde{b}_2$                 &  604  & 549\\  
\hline                                        
  $\tilde{t}_1$                 &  477  & 461\\  
\hline                                        
  $\tilde{t}_2$                 &  582  & 543\\  
\hline        \hline
  $ \tilde{\chi}^0_3(\tilde{H}_1)$&562  & (-)240\\ 
\hline
  $ \tilde{\chi}^0_4(\tilde{H}_2)$&(-)571& 248\\   
\hline
  $\tilde{\chi}^{\pm}_2(\tilde{H}^{\pm})$& 569& 254\\ 
\hline   \hline
  $       h $                   &  81   & 110\\ 
\hline
  $       H $                   &  739  & 273\\ 
\hline                                       
  $       A $                   &  734  & 273\\ 
\hline                                       
  $       H ^{\pm}$             &  738  & 285\\ 
\hline  \hline
  $\Omega h^2$                  &  0.42 & 0.19\\
\hline
  Br(\bsg)    & $2.87\;10^{-4}$ & $2.5\;10^{-4}$\\ 
\hline
  LSP$\rightarrow |\tilde{B}>$  & 0.9973 &  0.97\\  
\hline
  LSP$\rightarrow |\tilde{W}^3>$& 0.0360 & -0.05\\ 
\hline
  LSP$\rightarrow |\tilde{H}_1^0>$&-0.0593&-0.22\\ 
\hline
  LSP$\rightarrow |\tilde{H}_2^0>$& 0.0252&-0.03\\ 
\hline
\end{tabular} \end{center}
\caption[]{\label{t2}Values of the  SUSY  
            mass spectra for the low and high $\tb$ solutions, given in
            table \ref{t2a}. The (-) in front of the neutralinos indicates
            that it is a CP-odd state. The LSP is a linear combination of the
            gaugino and Higgsino components, as indicated by the last four rows.
            Note the much larger Higgsino component of the LSP for large
            $\tb$, which leads to a small relic density. }   
\end{table*}

\begin{table*}[tbh]
\begin{center}
\begin{minipage}{13cm}
\begin{minipage}[b]{5cm}
  %
  %
  \begin{center}
  $$
  \begin{array}{|c|c|c|c|}  \hline
  \mbox{Particle}  & b_1 & b_2 & b_3 \\
  \hline
  \hline
  \rule{0cm}{0.5cm}
  \tilde{g} & 0 & 0 & 2  \\
  \rule{0cm}{0.5cm}
  \tilde{l}_l & \frac{3}{10}  & \frac{1}{2} & 0  \\
  \rule{0cm}{0.5cm}
  \tilde{l}_r & \frac{3}{5}  &            0& 0  \\
  \rule{0cm}{0.5cm}
  \tilde{w}   &            0 &  \frac{4}{3}& 0  \\
  \rule{0cm}{0.5cm}
  \tilde{q}-\tilde{t} & \frac{49}{60}  &            1&\frac{5}{3}  \\
  \rule{0cm}{0.5cm}
  \tilde{t}_l         & \frac{1}{60}  & \frac{1}{2} &\frac{1}{6}  \\
  \rule{0cm}{0.5cm}
  \tilde{t}_r         & \frac{4}{15}  &           0 &\frac{1}{6}  \\
  \rule{0cm}{0.5cm}
  \tilde{h}           & \frac{2}{5}  & \frac{2}{3} &          0  \\
  \rule{0cm}{0.5cm}
             H        & \frac{1}{10}  & \frac{1}{6} &         0   \\
  \rule{0cm}{0.5cm}
         t            & \frac{17}{30}  &  1        &\frac{2}{3}  \\
                     & & & \\
  \hline
                     & & & \\
  \rule{0cm}{0.5cm}
   \mbox{SM}          & \frac{41}{10}  & -\frac{19}{6} & -7    \\
  \rule{0cm}{0.5cm}
   \mbox{MSSM}        & \frac{33}{5}  &            1  & -3    \\
                     & & & \\
  \hline
   \end{array}
$$
  \end{center}
   \caption[D]
           {\label{tbi}First order contributions to RG-equations.}
\end{minipage}\hfill
\begin{minipage}[b]{7cm}
 \begin{center}
 \renewcommand{\arraystretch}{1.545}
$$
 \begin{array}{|c|c|}
 \hline
  \mbox{Particle} & b_{ij}          \\
 \hline
 \hline
  \tilde{g} &
  \left(\begin{array}{rrr}
   \rule{0cm}{0.5cm}
               0&            0&    0        \\
   \rule{0cm}{0.5cm}
               0&            0&    0        \\
   \rule{0cm}{0.5cm}
              0      &      0      &    48
  \end{array}\right)   \\
 \hline
  \tilde{w} &
  \left(\begin{array}{rrr}
   \rule{0cm}{0.5cm}
               0      &      0      &     0 \\
   \rule{0cm}{0.5cm}
               0      & \frac{64}{3}&    0        \\
   \rule{0cm}{0.5cm}
               0      &      0      &     0
  \end{array}\right) \\
 \hline
  \tilde{q},\tilde{l},\mbox{gauginos} &
  \left(\begin{array}{rrr}
   \rule{0cm}{0.5cm}
   \frac{19}{15} & \frac{3}{5} & \frac{44}{15} \\
   \rule{0cm}{0.5cm}
   \frac{ 1}{5} &-\frac{7}{3} &       4       \\
   \rule{0cm}{0.5cm}
   \frac{11}{30} & \frac{3}{2} &-\frac{8}{3}
  \end{array}\right)   \\
 \hline
  \begin{array}{l} \mbox{Heavy Higgses} \\ \mbox{and Higgsinos}
  \end{array} &

  \left(\begin{array}{rrr}
   \rule{0cm}{0.5cm}
   \frac{9}{50} & \frac{9}{10} &    0 \\
   \rule{0cm}{0.5cm}
   \frac{3}{10}    & \frac{29}{6}&    0        \\
   \rule{0cm}{0.5cm}
               0      &      0      &     0
  \end{array}\right)   \\
 \hline
 \end{array} 
$$
 \renewcommand{\arraystretch}{1.0}
 \end{center}
  \caption[]
          {\label{tbij}Second order contributions to RG-equations.}
\end{minipage}
\end{minipage}
\end{center}
\end{table*}

%
%
\begin{table*}[htb]
\begin{center}
\renewcommand{\arraystretch}{1.2}
\begin{tabular}{|c|c|c|c|c|c|c|c|}
\hline
  Region I      &            & \multicolumn{6}{|c|}{ $\bf c^t_{ij}$}  \\
                              \cline{3-8}
  MSSM          & \rb{$\bf c^t_i$}
                & $\tal_1$            & $\tal_2$
                & $\tal_3$            & $Y_t$ 
                & $Y_b$               & $Y_\tau$                 \\
\hline
\hline
  $\tal_1$      & $ \frac{13}{15}$
                & $-\frac{2743}{450}$ & $-1$
                & $-\frac{136}{45}$   & $-\frac{6}{5}$  
                & $-\frac{2}{5}$      & $ 0$                        \\
  $\tal_2$      & $ 3$
                &                     & $-\frac{15}{2}$
                & $-8$                & $-6$
                & $ 0$                & $ 0$                        \\
  $\tal_3$      & $ \frac{16}{3}$     &        &
                & $ \frac{16}{9}$     & $-16$
                & $ 0$                & $ 0$                        \\
  $Y_t$         & $-6$                &        &
                &                     & $ 22$
                & $ 5$                & $  0$                       \\
  $Y_b$         & $-1$                &        &  &  &
                & $ 5$                & $  1$                       \\
  $Y_\tau$      & $ 0$                &        &  &  &  &  & $ 0$   \\
\hline
\hline
  Region II     &      &       &      &       &      &       &      \\
  SM            &      &       &      &       &      &       &      \\
\hline
\hline
  $\tal_1$      & $ \frac{17}{20}$
                & $-\frac{1187}{600}$ & $ \frac{9}{20}$
                & $-\frac{19}{15}$    & $-\frac{393}{80}$  
                & $-\frac{7}{80}$     & $-\frac{45}{24}$            \\
  $\tal_2$      & $ \frac{9}{4}$
                &                     & $ \frac{23}{4}$
                & $-9$                & $-\frac{225}{16}$
                & $-\frac{99}{16}$    & $-\frac{45}{24}$            \\
  $\tal_3$      & $ 8$                &        &
                & $ 108$              & $-36$
                & $-4$                & $ 0$                        \\
  $Y_t$         & $-\frac{9}{2}$      &        &
                &                     & $ 12$
                & $ \frac{11}{4}$     & $ \frac{9}{4}$              \\
  $Y_b$         & $-\frac{3}{2}$      &        &  &  &
                & $ \frac{1}{4}$      & $-\frac{5}{4}$              \\
  $Y_\tau$      & $-1$                &        &  &  &
                &                     & $ \frac{9}{4}$              \\
\hline
  \tabs{2cm}    & \tabs{1cm}
                & \tabs{1.1cm} & \tabs{1.1cm} & \tabs{1.1cm}
                & \tabs{1.1cm} & \tabs{1.1cm} & \tabs{1.1cm}
\end{tabular}
\renewcommand{\arraystretch}{1.}
\caption[]
  {\label{tctopij}One- and two-loop coefficients $c^t_i$ and
  $c^t_{ij}$ in the RG-equations.}
\end{center}
\end{table*}
%
%
\begin{table*}[htb]
\begin{center}
\renewcommand{\arraystretch}{1.2}
\begin{tabular}{|c|c|c|c|c|c|c|c|}
\hline
  Region I      &            & \multicolumn{6}{|c|}{ $\bf c^b_{ij}$}  \\
                              \cline{3-8}
  MSSM          & \rb{$\bf c^b_i$}
                & $\tal_1$            & $\tal_2$
                & $\tal_3$            & $Y_t$ 
                & $Y_b$               & $Y_\tau$                 \\
\hline
\hline
  $\tal_1$      & $ \frac{   7}{ 15}$
                & $-\frac{ 287}{ 90}$ & $-1$
                & $-\frac{   8}{  9}$ & $-\frac{  4}{ 5}$  
                & $-\frac{   2}{  5}$ & $-\frac{  6}{ 5}$          \\
  $\tal_2$      & $ 3$
                &                     & $-\frac{ 15}{ 2}$
                & $-8$                & $  0$
                & $-6$                & $  0$                       \\
  $\tal_3$      & $ \frac{  16}{  3}$ &        &
                & $ \frac{  16}{  9}$ & $  0$
                & $-16$               & $  0$                       \\
  $Y_t$         & $-1$                &        &
                &                     & $  5$
                & $ 5$                & $  0$                       \\
  $Y_b$         & $-6$                &        &  &  &
                & $ 22$               & $  3$                       \\
  $Y_\tau$      & $-1$                &        &  &  &  &  & $ 3$   \\
\hline
\hline
  Region II     &      &       &      &       &      &       &      \\
  SM            &      &       &      &       &      &       &      \\
\hline
\hline
  $\tal_1$      & $ \frac{   1}{ 4}$
                & $-\frac{ 127}{600}$ & $ \frac{ 27}{20}$
                & $-\frac{  31}{ 15}$ & $-\frac{ 91}{80}$  
                & $-\frac{ 237}{ 80}$ & $-\frac{ 15}{ 8}$           \\
  $\tal_2$      & $ \frac{   9}{  4}$
                &                     & $ \frac{ 23}{ 4}$
                & $-  9$              & $-\frac{ 99}{16}$
                & $-\frac{ 225}{ 16}$ & $-\frac{ 15}{ 8}$           \\
  $\tal_3$      & $   8$              &        &
                & $ 108$              & $- 4$
                & $- 36$              & $  0$                       \\
  $Y_t$         & $-\frac{   3}{  2}$ &        &
                &                     & $ \frac{  1}{ 4}$
                & $ \frac{  11}{  4}$ & $-\frac{  5}{ 4}$           \\
  $Y_b$         & $-\frac{   9}{  2}$ &        &  &  &
                & $  12$              & $ \frac{  9}{ 4}$           \\
  $Y_\tau$      & $-1$                &        &  &  &            
                &                     & $ \frac{  9}{ 4}$           \\
\hline
\hline
  Region III    &      &       &      &       &      &       &      \\
  SM - Top      &      &       &      &       &      &       &      \\
\hline
\hline
  $\tal_1$      & $ \frac{   1}{ 4}$
                & $-\frac{ 127}{600}$ & $ \frac{ 27}{20}$
                & $-\frac{  31}{ 15}$ & $  0$              
                & $-\frac{ 237}{ 80}$ & $-\frac{ 15}{ 8}$           \\
  $\tal_2$      & $ \frac{   9}{  4}$
                &                     & $ \frac{ 23}{ 4}$
                & $-  9$              & $  0$            
                & $-\frac{ 225}{ 16}$ & $-\frac{ 15}{ 8}$           \\
  $\tal_3$      & $   8$              &        &
                & $ 108$              & $  0$
                & $- 36$              & $  0$                       \\
  $Y_t$         & $   0$              &        &
                &                     & $  0$            
                & $   0$              & $  0$                       \\
  $Y_b$         & $-\frac{   9}{  2}$ &        &  &  &
                & $  12$              & $ \frac{  9}{ 4}$           \\
  $Y_\tau$      & $-1$                &        &  &  &
                &                     & $ \frac{  9}{ 4}$           \\
\hline
  \tabs{2cm}    & \tabs{1cm}
                & \tabs{1.1cm} & \tabs{1.1cm} & \tabs{1.1cm}
                & \tabs{1.1cm} & \tabs{1.1cm} & \tabs{1.1cm}
\end{tabular}
\renewcommand{\arraystretch}{1.}
\caption[E]
{\label{tcbotij}One- and two-loop coefficients $c^b_i$ and $c^b_{ij}$.
  The coefficients are given for three integration regions: $M_{\rm
    SUSY} < Q < M_{\rm GUT}$ (MSSM), $m_t < Q < M_{\rm SUSY}$ (SM) and
  $M_Z < Q < m_t$ (SM -- Top).  Below the top mass all coefficients
  $c^b_i$ and $c^b_{ij}$ proportional to $Y_t$ have to be set to
  zero.}
\end{center}
\end{table*}
%
%
\begin{table*}[htb]
\begin{center}
\renewcommand{\arraystretch}{1.2}
\begin{tabular}{|c|c|c|c|c|c|c|c|}
\hline
  Region I      &            & \multicolumn{6}{|c|}{ $\bf c^\tau_{ij}$}  \\
                              \cline{3-8}
  MSSM          & \rb{$\bf c^\tau_i$}
                & $\tal_1$            & $\tal_2$
                & $\tal_3$            & $Y_t$ 
                & $Y_b$               & $Y_\tau$                 \\
\hline
\hline
  $\tal_1$      & $ \frac{   9}{  5}$
                & $-\frac{  27}{  2}$ & $-\frac{  9}{ 5}$
                & $ 0$                & $  0$             
                & $ \frac{   2}{  5}$ & $-\frac{  6}{ 5}$          \\
  $\tal_2$      & $ 3$
                &                     & $-\frac{ 15}{ 2}$
                & $ 0$                & $  0$
                & $ 0$                & $- 6$                       \\
  $\tal_3$      & $ 0$                &        &
                & $ 0$                & $  0$
                & $-16$               & $  0$                       \\
  $Y_t$         & $ 0$                &        &
                &                     & $  0$
                & $ 3$                & $  0$                       \\
  $Y_b$         & $-3$                &        &  &  &
                & $ 9$                & $  9$                       \\
  $Y_\tau$      & $-4$                &        &  &  &  &  & $10$   \\
\hline
\hline
  Region II     &      &       &      &       &      &       &      \\
  SM            &      &       &      &       &      &       &      \\
\hline
\hline
  $\tal_1$      & $ \frac{   9}{ 4}$
                & $-\frac{1371}{200}$ & $-\frac{ 27}{20}$
                & $   0$              & $-\frac{ 17}{ 8}$  
                & $-\frac{   5}{  8}$ & $-\frac{537}{80}$           \\
  $\tal_2$      & $ \frac{   9}{  4}$
                &                     & $ \frac{ 23}{ 4}$
                & $   0$              & $-\frac{ 45}{ 8}$
                & $-\frac{  45}{  8}$ & $-\frac{165}{16}$           \\
  $\tal_3$      & $   0$              &        &
                & $   0$              & $-20$
                & $- 20$              & $  0$                       \\
  $Y_t$         & $-  3$              &        &
                &                     & $ \frac{ 27}{ 4}$
                & $-\frac{   3}{  2}$ & $ \frac{ 27}{ 4}$           \\
  $Y_b$         & $-  3$              &        &  &  &
                & $ \frac{  27}{  4}$ & $ \frac{ 27}{ 4}$           \\
  $Y_\tau$      & $-\frac{   5}{  2}$ &        &  &  &            
                &                     & $  3$                       \\
\hline
\hline
  Region III    &      &       &      &       &      &       &      \\
  SM - Top      &      &       &      &       &      &       &      \\
\hline
\hline
  $\tal_1$      & $ \frac{   9}{ 4}$
                & $-\frac{1371}{200}$ & $-\frac{ 27}{20}$
                & $   0$              & $  0$              
                & $-\frac{   5}{  8}$ & $-\frac{537}{80}$           \\
  $\tal_2$      & $ \frac{   9}{  4}$
                &                     & $ \frac{ 23}{ 4}$
                & $   0$              & $  0$            
                & $-\frac{  45}{  8}$ & $-\frac{165}{16}$           \\
  $\tal_3$      & $   0$              &        &
                & $   0$              & $  0$
                & $- 20$              & $  0$                       \\
  $Y_t$         & $-  3$              &        &
                &                     & $  0$            
                & $   0$              & $  0$                       \\
  $Y_b$         & $-  3$              &        &  &  &
                & $ \frac{  27}{  4}$ & $ \frac{ 27}{ 4}$           \\
  $Y_\tau$      & $-\frac{   5}{  2}$ &        &  &  &            
                &                     & $  3$                       \\
\hline
  \tabs{2cm}    & \tabs{1cm}
                & \tabs{1.1cm} & \tabs{1.1cm} & \tabs{1.1cm}
                & \tabs{1.1cm} & \tabs{1.1cm} & \tabs{1.1cm}
\end{tabular}
\renewcommand{\arraystretch}{1.}
\caption[]
{\label{tctauij} One- and two-loop coefficients $c^b_i$ and
  $c^b_{ij}$. The coefficients are given for three integration
  regions: $M_{\rm SUSY} < Q < M_{\rm GUT}$ (MSSM), $m_t < Q < M_{\rm
    SUSY}$ (SM) and $M_Z < Q < m_t$ (SM -- Top).  Below the top mass
  all coefficients $c^\tau_i$ and $c^\tau_{ij}$ proportional to $Y_t$
  have to be set to zero.}
\end{center}
\end{table*}
\clearpage

%
%
%
%
%
%
%
%
%
%
%
%
%
%
%
%
%
%
%
%
%
%
%
%
%
%
\clearpage
%
%
%
%
%
%
%
%
%
%
%
%
%
%
%
%
\clearpage
%
%
%
%
%
%
%
%
%

%
%
%
%

%
%
\clearpage

\clearpage
\addcontentsline{toc}{chapter}{References.} 
   \bibliographystyle{unsrt}
   \bibliography{biblio}

\end{document}